\documentclass[preprint,aps,prc,showpacs,nofootinbib]{revtex4-1}
\usepackage{amsmath}
\usepackage{amssymb}
\usepackage{graphics}
\usepackage{epsfig}

\usepackage{subfigure}

\newcommand\nn{\nonumber}
\newcommand\ba{\begin{eqnarray}}
\newcommand\ea{\end{eqnarray}}
\newcommand\be{\begin{equation}}
\newcommand\ee{\end{equation}}

\begin{document}

\title{General analysis of polarization effects in coherent
pseudoscalar mesons photo-- and electroproduction on deuteron}

\author{G. I. Gakh} 
\affiliation{\it National Science Centre, Kharkov Institute of Physics and Technology, \\ 61108 Akademicheskaya 1, Kharkov, and V. N. Karazin Kharkov National University, Dept. of
Physics and Technology, 31 Kurchatov, 61108, Kharkov, Ukraine}
\author{A. G. Gakh}
\affiliation{\it V. N. Karazin Kharkov National University, Dept. of
Physics and Technology, 31 Kurchatov, 61108, Kharkov, Ukraine}

\author{E. Tomasi--Gustafsson}
\email{egle.tomasi@cea.fr}
\affiliation{\it IRFU, CEA, Universit\'e Paris-Saclay, 91191 Gif-sur-Yvette Cedex, France}

\begin{abstract}
General expressions for the unpolarized differential cross section and for
various polarization observables in the coherent pseudoscalar meson
photo- and electroproduction on the deuteron target have been
obtained in one-photon-exchange approximation. The spin structure of the matrix element is explicitly derived in terms of structure functions. 
The correspondence with the helicity amplitudes is given.
The polarization
effects have been investigated for the case of a longitudinally
polarized electron beam and vector or tensor polarized deuteron target. 
The polarization (vector or tensor) of the scattered deuteron for the case of unpolarized or a
longitudinally polarized electron beam is also considered.
In the case of the photoproduction
reaction, we consider linearly, circularly or elliptically polarized
photon beam. 
Numerical
estimations have been done for the
unpolarized differential cross section and for some polarization
observables.

\end{abstract}
\maketitle

\section{Introduction}
\label{Section:Introduction}

The complete characterization of meson photo- and electroproduction
on nuclei requires the detection of the neutron which
must rely on meson  production by real or virtual photons from light
nuclei. The well known nuclear
structure of the deuteron makes it a most suitable nuclear target, in comparison to other nuclei.
Experiments with deuteron targets have been done and are being
performed. A recent status of the study of the meson 
photoproduction can be found in the review \cite{Krusche:2003ik}.

Besides the experiments on the electromagnetic production of the
isovector mesons ($\pi $-meson), a series of precise measurements of
breakup and coherent isoscalar meson ($\eta $-meson) photoproduction
from the deuteron target have been performed during the last few
years. The inclusive cross section of the reactions $d(\gamma ,\eta
)X$ are presented in Refs. \cite{Krusche:1995zx,Weiss:2002tn}. 
 The exclusive
reaction with detection of the recoil nucleons was also investigated
for the deuteron target \cite{Weiss:2002tn,HoffmannRothe:1997sv,Hejny:1999iw}. The data on coherent
$\eta $-meson photoproduction off the deuteron are given in Refs.
\cite{HoffmannRothe:1997sv,Weiss:2001yy}.

The combination of $4\pi $ detectors (as it can be recently done in Bonn and
Mainz) with linearly and circularly polarized photon beams as well
as polarized targets will provide the measurement of various
polarization observables. These new observables can give us an
additional information on resonance properties and the details of
the reaction mechanism.

In the last decade the progress in the investigation of the meson
production by the electromagnetic probes has been substantial and we
moved forward in the understanding of the resonance properties.

The differential cross sections for the reaction on coherent and
incoherent $\pi^0$-meson photoproduction from the deuteron have been
measured at MAMI (Mainz) in the energy range $140$~ MeV $< E_{\gamma}
< 306$ MeV \cite{Siodlaczek:2001mh}. Earlier, the total and differential cross
sections covering the full angular range have been obtained for
coherent and incoherent single $\pi^0$-meson photoproduction from
the deuteron in the second resonance region (200~ MeV $< E_{\gamma}
< 792$ MeV) at Mainz \cite{Krusche:1999tv}. It was found that final-state
interaction effects for the incoherent process are much more
important than for coherent reaction. In the last case these effects
are not very large and different models do not agree about the main
final-state interaction mechanism. Therefore,  no final conclusion
about these mechanisms can be reached.

The data on pion production in coherent electron-deuteron collisions are
scarce. The experimental study of this reaction is now possible, at
Mainz and JLab, due to the high-duty cycle of the electron machines.
Threshold $\pi^0$-meson electroproduction on protons and deuteron
has been investigated by the A1 collaboration at Mainz \cite{Distler:1998ae,Ewald:2000sx}
at small four momentum transfer squared $Q^2\le 0.1$~ GeV$^2$. The first experimental results for the
coherent $\pi^0$-meson electroproduction off the deuteron at large
$Q^2$, $1.1 < Q^2 < 1.8$~ GeV$^2$, from the threshold to 200 MeV
excitation energy in the $d\pi^0$ system, are reported in Ref.
\cite{TomasiGustafsson:2003nj}. The data were collected during the $t_{20}$ experiment,
the primary aim of which was the measurement of the deuteron tensor
polarization in elastic electron-deuteron scattering \cite{Abbott:2000fg}.

A general theoretical study of pion electroproduction on deuterons
was first developed in Ref. \cite{Rekalo:2002km} for the unpolarized case. The reaction $e+d\to
e+d+\pi^{0}$ involves the study of the deuteron structure and of the
reaction mechanism, and requires the knowledge of the neutron and proton elementary
amplitudes, $\gamma^*+n,p\to n,p+\pi^{0}$.

The experimental investigation of the nucleon resonance properties
by means of the meson production processes can be used to verify and constrain 
the models of the hadron structure. The
production of the neutral mesons by  real or virtual photons is
of special interest since,in these
reactions, the background contributions  are suppressed due to the weak coupling of the photon with
neutral mesons.

In this work, we follow the formalism of Ref.  \cite{Rekalo:2002km}. 
We focus on the polarization observables, giving the general expressions  for the case of a longitudinally polarized electron beam and 
vector or tensor polarized deuteron target (or scattered deuteron), in the one-photon exchange approximation and neglecting the lepton mass. In the case of the photoproduction reaction, we consider linearly, circularly or elliptically polarized photon beam. A numerical application and illustration of the
observables is shown, on the basis of impulse approximation and the elementary model previously developed in Ref. \cite{Rekalo:2002km}. 

Not only high intensity polarized electron beams are available, and vector and tensor polarized targets are currently used, but also the principle of measuring the vector and tensor deuteron polarization in the GeV range from backward elastic scattering,  has been proved to be feasible \cite{TomasiGustafsson:1998wv}.

\section{Matrix element and differential cross section}

The general structure of the differential cross section for the reaction
\be
\label{eq:eqreac}
e^-(k_1)+d(p_1)\to e^-(k_2)+d(p_2)+\pi^0(q),
\ee
(the four--momenta of the corresponding particles are indicated in the
brackets)
can be written in the frame of the one--photon--exchange mechanism. 
The formalism in this section is based on
the most general symmetry properties of the hadron electromagnetic
interaction, such as the gauge invariance (the conservation of the hadronic
and leptonic electromagnetic currents), $P$ --invariance (the invariance with
respect to the space  reflections) and does not depend on the deuteron structure
and on the details of the reaction mechanism. 

In the one--photon--exchange approximation, the matrix element for
the process of the coherent $\pi^0$--meson electroproduction on the
deuteron can be written  as
\be
M(ed\to ed \pi^0)=\frac{e^2}{k^2}j_{\mu}J_{\mu},
\label{eq:MatrixElement}
\ee
with
\be
j_{\mu}=\bar u(k_2)\gamma _{\mu}u(k_1), \ \
J_{\mu}=<d\pi^0|\hat J_{\mu}|d>, 
\label{eq:MatrixEl2}
\ee
where $k=k_1-k_2$ is the virtual--photon four--momentum and $J_{\mu}$ is the
electromagnetic current describing the transition $\gamma ^*+d\to d+\pi^0$
($\gamma ^*$ is the virtual photon).

The electromagnetic structure of nuclei, as probed by elastic and inelastic
electron scattering by nuclei, can be characterized by a set of response
functions or structure functions \cite{Donnelly:1985ry,Boffi:1996}. Each of these structure functions
is determined by different combinations of the longitudinal and transverse
components of the electromagnetic current $J_{\mu}$, thus providing different
pieces of information about the nuclear structure or possible mechanisms of
the reaction under consideration. The components which are determined by the
real parts of the bilinear combinations of the reaction amplitudes, are nonzero
in the impulse approximation, the other ones, which originate from the imaginary
part of structure functions, vanish if the final state interaction is absent.

The formalism of the structure functions is especially convenient for the
investigation of polarization phenomena for the reaction (\ref{eq:eqreac}).

Using the conservation of the leptonic $j_{\mu}$ and hadronic $J_{\mu}$
electromagnetic currents $(k\cdot j=k\cdot J=0)$, one can rewrite the matrix
element (\ref{eq:MatrixElement}) in terms of the space components of these currents only
\be
M(ed\to ed\pi^0)=\frac{e^2}{k^2}{\vec e}\cdot {\vec J}, \ \
{\vec e}=\frac{{\vec j}\cdot {\vec k}}
{k_0^2}{\vec k}-{\vec j},
\label{Eq:Matrix3}
\ee
where $k=(k_0, {\vec k})$ and $k_0({\vec k})$ is the energy (three--momentum)
of the virtual photon in CMS of the $\gamma ^*+d\to d+\pi^0$ reaction.
All observables are determined by bilinear combinations of the space
components of the hadronic current ${\vec J}$: $H_{ab}=J_aJ_b^*.$ As a result,
we obtain the following general structure of the differential cross section
for the reaction (\ref{eq:eqreac}), when the scattered electron and $\pi^0$--meson are
detected in coincidence, and the electron beam is longitudinally polarized
(the polarization states of the deuteron target and scattered deuteron can
be any)
\ba
\frac{d^3\sigma}{dE'd\Omega_ed\Omega_{P}} &=&
N\biggl [H_{xx}+H_{yy}+\varepsilon \cos(2\varphi)(H_{xx}-H_{yy})+
\varepsilon \sin(2\varphi)(H_{xy}+H_{yx})\nn\\
&&-2\varepsilon\frac{k^2}{k_0^2}H_{zz}-
\frac{\sqrt{-k^2}}{k_0}\sqrt{2\varepsilon (1+\varepsilon)}
\cos\varphi(H_{xz}+H_{zx})-\nn\\
&&
-\frac{\sqrt{-k^2}}{k_0}\sqrt{2\varepsilon (1+\varepsilon)}
\sin\varphi(H_{yz}+H_{zy}) \nn\\
&&
\mp i\lambda\sqrt{(1-\varepsilon ^2)}(H_{xy}-H_{yx})\mp
i\lambda \frac{\sqrt{-k^2}}{k_0}\sqrt{2\varepsilon (1-\varepsilon)}
\cos\varphi(H_{yz}-H_{zy})\pm 
\nn\\
&&
\pm i\lambda \frac{\sqrt{-k^2}}{k_0}\sqrt{2\varepsilon (1-\varepsilon)}
\varphi(H_{xz}-H_{zx})\biggr ], \nn\\
&&
N=\frac{\alpha^2}{64\pi^3}\frac{E'}{E}\frac{|{\vec q}|}{MW}
\frac{1}{1-\varepsilon}\frac{1}{(-k^2)}, 
|{\vec k}|=\frac{1}{2W}\sqrt{(W^2+M^2-k^2)^2-4M^2W^2},\nn\\
&&
|{\vec q}|=\frac{1}{2W}\sqrt{(W^2+M_{P}^2-M^2)^2-4M_{P}^2W^2}, 
\ \varepsilon^{-1}=1-2\frac{{\vec k}^2_{Lab}}{k^2}\tan^2(\frac{\theta_e}{2}).
\label{Eq:sigma}
\ea
The $z$ axis is directed along the virtual photon momentum ${\vec
k}$, the momentum of the detected $P$--meson ${\vec q}$ lies in the
$xz$ plane (reaction plane); $E (E')$ is the energy of the initial
(scattered) electron in the deuteron rest frame (laboratory (Lab)
system); $\theta_e $ is the electron scattering angle (the angle
between the momenta of the initial and scattered electrons) in the
Lab system; $d\Omega_e$ is the solid angle of the scattered electron
in the Lab system, $d\Omega_{P}(q)$ is the solid angle (value of the
three-momentum) of the detected $P$--meson in the $Pd$--pair
center--of--mass system (CMS), $M_{P}, M$ are the masses of the
$P$--meson, deuteron, respectively; $\varphi$ is the azimuthal angle
between the electron scattering plane and the plane where the
detected $P$--meson lies $(xz)$, $k_0=(W^2+k^2-M^2)/2W$ is the
virtual photon energy in the $Pd$--pair CMS, $W$ is the invariant
mass of the final hadrons, $W^2=M^2+k^2+2M(E-E')$; $\lambda$ is the
degree of the electron longitudinal polarization, $\varepsilon $ is
the degree of the linear polarization of the virtual photon. The
upper (bottom) sign in this formula corresponds to the electron
(positron) scattering. This expression is valid for zero electron
mass. Below we will neglect it wherever possible.

Let us introduce, for convenience and simplification of the following
calculations of the polarization observables, the orthonormal system
of basic unit vectors ${\vec m}, {\vec n}$, and $\hat {\vec k}$
which are built from the momenta of the particles participating in
the reaction under consideration
$$\hat {\vec k}=\frac{{\vec k}}{|{\vec k}|}, \ \
{\vec n}=\frac{{\vec k}\times{\vec q}}{|{\vec k}\times{\vec q}|}, \ \
{\vec m}={\vec n}\times\hat {\vec k}. $$
The unit vectors $\hat {\vec k}$ and ${\vec m}$ define the $\gamma^* +
d\to d+\pi^0$ reaction $xz$--plane ($z$ axis is directed along the three--momentum
of the virtual photon ${\vec k}$, and $x$ axis is directed along the unit
vector ${\vec m}$), and the unit vector ${\vec n}$ is perpendicular to
the reaction plane.

First of all, let us establish the spin structure of the matrix element for
the $\gamma^* +d\to d+\pi^0$ reaction without any constraint on the kinematical
conditions.

The amplitude spin structure can be parameterized by different (and
equivalent) methods, but for the analysis of the polarization
phenomena the choice of the transverse amplitudes is sometimes
preferable. Taking into account the $P$--invariance of the hadron
electromagnetic interaction, the dependence of the $\gamma^* +d\to
d+\pi^0$ amplitude on the virtual--photon polarization vector and
polarization three vectors ${\vec U}_1$ and  ${\vec U}_2$ of the
initial and final deuterons is given by \cite{Rekalo:2002km}:
\ba
F(\gamma^* d\to dP)&=&{\vec e}\cdot {\vec m}(
g_1{\vec m}\cdot {\vec U}_1{\vec n}\cdot {\vec U}_2^*+
g_2\hat {\vec k}\cdot {\vec U}_1{\vec n}\cdot {\vec U}_2^*+
g_3{\vec n}\cdot {\vec U}_1{\vec m}\cdot {\vec U}_2^*+
g_4{\vec n}\cdot {\vec U}_1\hat {\vec k}\cdot {\vec U}_2^*)+
\nn\\ 
&&
+{\vec e}\cdot {\vec n}\bigl(
g_5{\vec m}\cdot {\vec U}_1{\vec m}\cdot {\vec U}_2^*+
g_6{\vec n}\cdot {\vec U}_1{\vec n}\cdot {\vec U}_2^*+
g_7\hat {\vec k}\cdot {\vec U}_1\hat {\vec k}\cdot {\vec U}_2^*+
\nn\\ 
&&
+g_8{\vec m}\cdot {\vec U}_1\hat {\vec k}\cdot {\vec U}_2^*
g_9\hat {\vec k}\cdot {\vec U}_1{\vec m}\cdot {\vec U}_2^*)+ 
{\vec e}\cdot \hat {\vec k}(
g_{10}{\vec m}\cdot {\vec U}_1{\vec n}\cdot {\vec U}_2^*+
\nn\\ 
&&
+g_{11}\hat {\vec k}\cdot {\vec U}_1{\vec n}\cdot {\vec U}_2^*+
g_{12}{\vec n}\cdot {\vec U}_1{\vec m}\cdot {\vec U}_2^*+
g_{13}{\vec n}\cdot {\vec U}_1\hat {\vec k}\cdot {\vec U}_2^*\bigr ), 
\label{eq:eqampl}
\ea
where $g_i (i=1-13)$ are the scalar amplitudes, depending on three variables
$k^2$, $W$ and $\vartheta $ ($\vartheta $ is the angle between the virtual
photon and $\pi^0$--meson momenta in the $\gamma^* +d\to d+\pi^0$ reaction CMS),
which completely determine the reaction dynamics. If we single out the
virtual--photon polarization vector ${\vec e}$, we can write the
amplitude ${\it F}$ as follows
$${\it F}={\it F_i}e_i $$
and the hadronic tensor can be written  in terms of ${\it F_i}$ as
$$H_{ij}={\it F_i}{\it F_j^*}. $$

The process $\gamma^* +d\to d+P$ is described by
a set of nine amplitudes for the absorption of a virtual photon with
transverse polarization and four amplitudes for the absorption of a
virtual photon with longitudinal polarization. These numbers are
dictated by the values of the spins of the particles  and by the
$P$--invariance of hadron electrodynamics. Therefore, the complete
experiment requires, at least, the measurement of 25 observables.
Let  us mention in this respect specific properties of polarization
phenomena for inelastic electron--hadron scattering: in exclusive
$e^- +d\to e^-+d+P$ processes the virtual photon has a nonzero
linear polarization, even for the scattering of unpolarized
electrons by an unpolarized deuteron target.

\section{Polarized deuteron target}

Let us consider the dependence of the observables on the polarization state of the deuteron
target, which is described by the spin density matrix. We use the following
general expression for the deuteron spin density matrix in the coordinate
representation \cite{Schildknecht:1965,Schildknecht:1967}
\be
\rho_{\mu\nu}=-\frac{1}{3}(g_{\mu\nu}-\frac{p_{1\mu}p_{1\nu}}{M^2})-
\frac{i}{2M}\varepsilon _{\mu\nu\alpha\beta}s_{\alpha}p_{1\beta}+S_{\mu\nu},
\label{eq:eqdm}
\ee
where $s_{\alpha}$ is the four--vector describing the vector polarization of
the target, $s^2=-1,$ $s\cdot p_1=0$ and $S_{\mu\nu}$ is the tensor describing
the tensor (quadrupole) polarization of the target, $S_{\mu\nu}=S_{\nu\mu},$
$p_{1\mu}S_{\mu\nu}=0, $ $S_{\mu\mu}=0 $ (due to these properties the tensor
$S_{\mu\nu}$ has only five independent components). In Lab system all time
components of the tensor $S_{\mu\nu}$ are zero and the tensor polarization
of the target is described by five independent space components $(S_{ij}=
S_{ji}, S_{ii}=0, i,j=x,y,z).$ The four--vector $s_{\alpha}$ is related to
the unit vector ${\vec \xi}$ of the deuteron vector polarization in its rest
system by:
 \be
s_0=-{\vec k}{\vec \xi}/M, \ {\vec s}={\vec \xi}+
{\vec k}({\vec k}{\vec \xi})/M(M+E_1),
\label{eq:pol}
\ee 
where $E_1$ is the deuteron--target energy
in the $\gamma ^*+d\to d+\pi^0$ reaction CMS.

The hadronic tensor $H_{ij} (i,j=x,y,z)$ depends linearly on the target
polarization and  can be written as 
\be
H_{ij}=H_{ij}(0)+H_{ij}(\xi)+H_{ij}(S),
\label{eq:hadrontensor}
\ee
where the term $H_{ij}(0)$ corresponds to the case of  unpolarized deuteron
target, and the term $H_{ij}(\xi) (H_{ij}(S))$ corresponds to the case of 
vector (tensor)-polarized target.
\subsection{Unpolarized deuteron target}
The general structure of the part of the hadronic tensor which corresponds to
the unpolarized deuteron target has following form:
\be
H_{ij}(0)=h_1m_im_j+h_2n_in_j+h_3\hat k_i\hat k_j+h_4\{m,\hat k\}_{ij}+
ih_5[m,\hat k]_{ij},
\label{eq:8}
\ee
where $\{a,b\}_{ij}=a_ib_j+a_jb_i, \ [a,b]_{ij}=a_ib_j-a_jb_i $ and
the real structure functions $h_i$ depend on three invariant
variables $s=W^2=(k+p_1)^2$, $k^2$ and $t=(k-p_1)^2$. The structure
functions $h_1-h_4$ determine the cross section for the
$e^-+d\to e^-+d+P$ reaction with unpolarized particles. Let
us emphasize that the structure function $h_5$ (the so--called fifth
structure function) determines the asymmetry of longitudinally
polarized electrons scattered by an unpolarized target. It is
determined by the strong interaction effects of the $P$--meson and
deuteron in the final state  and it vanishes for the pole (Born)
diagram contribution in all kinematic range (independently on the
particular parametrization of the $\gamma^* N\to NP$ amplitude and
$dnp-$vertex). This is true for the nonrelativistic approach and for
the relativistic one as well describing the $\gamma^*
+d\to d+P$ reaction. The scattering of longitudinally
polarized electrons by unpolarized deuteron target allows to
determine the $h_5$ contribution. Then the corresponding asymmetry
is determined only by the strong interaction effects. More exactly,
it is determined by the effects arising from non-pole mechanisms of
various nature: meson exchange currents can  induce nonzero
asymmetry,  dibaryon resonances, if they exist, lead also to
nonzero asymmetry.

In the chosen coordinate system, the different hadronic tensor components,
entering in the expression of the cross section (4), are related to the
structure functions $h_i (i=1-5)$ by:
\ba H_{xx}\pm H_{yy}&=&h_1\pm h_2, \ \  H_{zz}=h_3, \ \
H_{xz}+ H_{zx}=2h_4, \nn\\
H_{xz}- H_{zx}&=&2ih_5, \ \  H_{xy}\pm H_{yx}=0,  \ \
H_{yz}\pm H_{zy}=0. 
\label{eq:hh}
\ea
The expressions for the structure functions
$h_i$ (i=1-5) in terms of the reaction amplitudes $g_i (i=1-13)$ are
given in  Appendix I. The expressions of the reaction amplitudes $g_i (i=1-13)$ 
depend on the underlying model. Their explicit form as function of the deuteron inelastic form factors in impulse approximation can be found in Ref. \cite{Rekalo:2002km}.

In the one--photon--exchange approximation, the general structure of the
differential cross section for the reaction $d({\vec e}, e'P)d$ (in the
case of longitudinally polarized electron beam and unpolarized deuteron
target) can be written in terms of five independent contributions
\be\label{definition of cross--section, 9}
\frac{d^3\sigma}{dE'd\Omega_ed\Omega_{P}} =
N\biggl [\sigma_{T}+\varepsilon \sigma_{L}+
\varepsilon \cos(2\varphi)\sigma_{P}+
\sqrt{2\varepsilon (1+\varepsilon )} \cos\varphi \sigma_{I}+
\lambda \sqrt{2\varepsilon (1-\varepsilon )}\varphi \sigma'_{I}
\biggr ],
\ee
where the individual contributions are related to the structure
functions $h_i$ of the spin--independent hadronic tensor, Eq. (\ref{eq:8}),
by:
\be
\sigma_{T}=h_1+h_2, ~ \sigma_{P}=h_1-h_2, ~
\sigma_{L}=-2\frac{k^2}{k_0^2}h_3, 
\sigma_{I}=-2\frac{\sqrt{-k^2}}{k_0}h_4, ~
\sigma'_{I}=-2\frac{\sqrt{-k^2}}{k_0}h_5.
\label{eq:10}
\ee
One can see from this equation that it exists a single--spin
asymmetry due to the longitudinal polarization of the electron
beam and it is defined as:
\begin{equation}\label{11}
\Sigma_e(\varphi )=\frac{d\sigma (\lambda =+1)-d\sigma (\lambda
=-1)} {d\sigma (\lambda =+1)+d\sigma (\lambda =-1)}=
\end{equation}
$$=\frac{sin\varphi \sqrt{2\varepsilon (1-\varepsilon )}\sigma'_{I}}
{\sigma_{T}+\varepsilon \sigma_{L}+\varepsilon \cos(2\varphi)\sigma_{P}+
\sqrt{2\varepsilon (1+\varepsilon )}\cos\varphi \sigma_{I}}. $$
Due to the $\varphi $--dependence, this asymmetry has to be measured in
noncoplanar geometry (out--of--plane kinematics).

For the case of unpolarized particles, one can determine the
so called left--right asymmetry
\be
\label{12}
A_{LR}=\frac{d\sigma (\varphi =0^0)-d\sigma (\varphi =180^0)}
{d\sigma (\varphi =0^0)+d\sigma (\varphi
=180^0)}=\frac{\sqrt{2\varepsilon (1+\varepsilon )}\sigma_{I}}
{\sigma_{T}+\varepsilon ( \sigma_{L}+\sigma_{P})}.
\ee
We see that the $\Sigma_e(\varphi )$ asymmetry is determined by the
structure function $h_5$ which is defined by the interference of the
reaction amplitudes characterizing the absorption of virtual
photons with nonzero longitudinal and transverse components of the
electromagnetic current corresponding to the process 
$\gamma^*+d\to d+\pi^0$. One finds that $h_5\sim \sin\vartheta $
($\vartheta $ is the angle between three--momenta of the virtual
photon and the P-meson in the CMS of the $\gamma ^*+d\to d+\pi^0$ reaction)
for any reaction mechanism of the considered reaction. It
vanishes in collinear kinematics, i.e., at $\pi^0$--meson emission angles $\vartheta =0^{\circ}$ and
$\vartheta =180^{\circ}$ due to the conservation of the total
helicity of the interacting particles. The structure function $h_5$ is nonzero only if the
complex amplitudes of the $\gamma ^*+d\to d+P$ reaction have
nonzero relative phases. This is a very specific observable, which
has no counterpart in the  process of the $P$--meson
photoproduction on the deuteron $\gamma +d\to d+\pi^0$.

The study of the single--spin asymmetry $\Sigma_e$ was firstly suggested for
the pion production in the electron--nucleon scattering, $e+N\to e+N+\pi $
\cite{Gehlen1971141}. Afterwards this asymmetry has been dicussed for the hadron
production in the exclusive processes of the type $A({\vec e}, e'h)X$,
where $A$ is a nucleus and $h$ is the detected hadron \cite{Boffi:1985pc,PhysRevC.32.1312}. A number
of experiments have  measured the asymmetry $\Sigma_e$ \cite{PhysRevLett.72.3325,PhysRevLett.88.142001,PhysRevC.51.3479}.
\subsection{Vector polarized deuteron target}

The part of the hadronic tensor depending on the deuteron vector polarization
has the following general structure:
\ba
H_{ij}(\xi )&=&{\vec\xi }{\vec m}(h_{6}\{m,n\}_{ij}+
h_{7}\{\hat k,n\}_{ij}+ih_{8}[m,n]_{ij}+ih_{9}[\hat k,n]_{ij})+
\nn\\ 
&&
+{\vec\xi }{\vec n}(h_{10}m_im_j+h_{11}n_in_j+h_{12}\hat k_i\hat k_j+
h_{13}\{m,\hat k\}_{ij}+ih_{14}[m,\hat k]_{ij})+ \nn\\ 
&&
+{\vec\xi }\hat {\vec k}(h_{15}\{m,n\}_{ij}+h_{16}\{\hat k,n\}_{ij}+
ih_{17}[m,n]_{ij}+ih_{18}[\hat k,n]_{ij}).
\label{eq:vecpol}
\ea
where one can see that , the dependence of the polarization observables on the
deuteron vector polarization is determined by 13 structure
functions. The expressions for these structure functions in terms of
the reaction amplitudes $g_i$, $(i=1-13)$ are given in  Appendix I. On
the basis of this formula one can make the following general
conclusions:
\begin{enumerate}
\item
If the deuteron is vector polarized and the polarization vector 
is perpendicular to the $\gamma ^* +d\to d+P$ reaction plane,
then the dependence of the differential cross section of the 
$e^-+d \to e^-+d+P$ reaction on the $\varepsilon$ and $\varphi$ variables is the
same as in the case of the unpolarized target, and the nonvanishing components
of the $H_{ij}(\xi )$ tensor are:
\ba
H_{xx}(\xi ) \pm H_{yy}(\xi )&=&(h_{10}\pm h_{11}){\vec\xi }{\vec n},
 \ \ H_{zz}(\xi )=h_{12}{\vec\xi }{\vec n}, \nn\\
H_{xz}(\xi )+H_{zx}(\xi )&=&2h_{13}{\vec\xi }{\vec n}, \ \
H_{xz}(\xi )-H_{zx}(\xi )=2ih_{14}{\vec\xi }{\vec n}. 
\label{eq:hxyz}
\ea
\item
If the deuteron target is polarized in the $\gamma ^* +d\to d+\pi^0$
reaction plane (in the direction of the vector ${\vec k}$ or ${\vec
m}$), then the dependence of the differential cross section of the
$e^-+d\to e^-+d+\pi^0$ reaction on the $\varepsilon$ and $\varphi$
variables is:
\begin{itemize}
\item
  for deuteron P-meson production by an unpolarized electron beam:
\be
\varepsilon \sin(2\varphi), \ \ \sqrt{2\varepsilon
(1+\varepsilon )}\sin\varphi,
\label{eq:e1}
\ee
\item for deuteron P-meson production by a longitudinally polarized
electron beam:
\be
\pm i\lambda\sqrt{1-\varepsilon ^2}, \ \ \mp i\lambda\sqrt{2\varepsilon
(1-\varepsilon )}\cos\varphi. 
\label{eq:e2}
\ee
\end{itemize}
\item
The differential cross section of the reaction ${\vec d}({\vec
e},e'P)d$, where the electron beam is  longitudinally polarized and
the deuteron target is  vector polarized, can be written as
follows:
\be
\label{definition of cross--section}
\frac{d^3\sigma}{dE'd\Omega_e d\Omega_{P}} =
\sigma_0\biggl [1+\lambda \Sigma_{e}+(A_x^d+\lambda A_x^{ed})\xi_x+
(A_y^d+\lambda A_y^{ed})\xi_y+(A_z^d+\lambda A_z^{ed})\xi_z
\biggr ],
\ee
where $\sigma_0$ coincides with the five-fold  unpolarized differential cross section, Eq. (\ref{Eq:sigma}), 
$\Sigma _{e}$ is the beam analyzing power (the asymmetry induced by the electron--beam
polarization), $A_i^d (i=x,y,z)$ are the analyzing powers due to the vector
polarization of the deuteron target, and $A_i^{ed}$, $(i=x,y,z)$ are the 
spin--correlation parameters. The direction of the deuteron polarization vector is
defined by the angles $\vartheta^*$, $\varphi^*$ in the frame where the $z$
axis is along the direction of the three--momentum transfer ${\vec k}$ and
the $y$ axis is defined by the vector product of the detected $\pi^0$-- meson
and virtual photon momenta (along the unit vector ${\vec n}$). The
target analyzing powers and spin--correlation parameters depend on the
orientation of the deuteron polarization vector. The quantities $\Sigma_{e}$
and $A_i^d$ are T--odd observables and they are entirely determined by the
reaction mechanisms beyond the impulse approximation, for example, by 
final--state interaction effects. On the contrary, the quantities $A_i^{ed}$
are T--even observables and they do not vanish even in the absence of 
final--state interaction effects.

The expressions of the $A_i^{d}$ and $A_i^{ed}$ asymmetries can be explicitly
written as functions of the azimuthal angle $\varphi $, of the virtual--photon
linear polarization $\varepsilon $, and of contributions of the longitudinal
(L) and transverse (T) components (relative to the virtual--photon momentum
${\vec k}$) of the hadron electromagnetic current of  $\gamma^*+d\to
d+\pi^0$ :
\ba
A_x^d\sigma_0&=&N\sin\varphi \biggl [\sqrt{2\varepsilon (1+\varepsilon )}
A_x^{(LT)}+\varepsilon \cos\varphi A_x^{(TT)}\biggr ], \nn\\
A_z^d\sigma_0&=&N\sin\varphi \biggl [\sqrt{2\varepsilon (1+\varepsilon )}
A_z^{(LT)}+\varepsilon \cos\varphi A_z^{(TT)}\biggr ], \nn\\
A_y^d\sigma_0&=&N\biggl [A_y^{(TT)}+\varepsilon A_y^{(LL)}+
\sqrt{2\varepsilon (1+\varepsilon )}cos\varphi A_y^{(LT)}+
\varepsilon cos(2\varphi )\bar A_y^{(TT)}\biggr ], \nn\\
A_x^{ed}\sigma_0&=&N\biggl [\sqrt{1-\varepsilon^2}
B_x^{(TT)}+\sqrt{2\varepsilon (1-\varepsilon )}
cos\varphi B_x^{(LT)}\biggr ], \nn\\
A_z^{ed}\sigma_0&=&N\biggl [\sqrt{1-\varepsilon^2}
B_z^{(TT)}+\sqrt{2\varepsilon (1-\varepsilon )}
cos\varphi B_z^{(LT)}\biggr ], \nn\\
A_y^{ed}\sigma_0&=&N\sqrt{2\varepsilon (1-\varepsilon )}
\sin\varphi B_y^{(LT)}, 
\label{eq:34}
\ea 
where $N$ is defined in Eq. (\ref{Eq:sigma}) and 
the individual contributions to the considered asymmetries in terms of
the structure functions $h_i$ are given by
\ba
A_x^{(TT)}&=&4h_{6}, \ \ A_y^{(TT)}=h_{10}+h_{11}, \ \
\bar A_y^{(TT)}=h_{10}-h_{11}, \ \
A_z^{(TT)}=4h_{15},  \nn\\
A_x^{(LT)}&=&-2\frac{\sqrt{Q^2}}{k_0}h_{7}, \ \
A_y^{(LT)}=-2\frac{\sqrt{Q^2}}{k_0}h_{13}, \ \
A_z^{(LT)}=-2\frac{\sqrt{Q^2}}{k_0}h_{16},  \nn\\
A_y^{(LL)}&=&2\frac{Q^2}{k_0^2}h_{12}, \ \
B_x^{(TT)}=2h_{8}, \ \ B_z^{(TT)}=2h_{17},  \nn\\
B_x^{(LT)}&=&-2\frac{\sqrt{Q^2}}{k_0}h_{9}, \ \
B_y^{(LT)}=-2\frac{\sqrt{Q^2}}{k_0}h_{14}, \ \
B_z^{(LT)}=-2\frac{\sqrt{Q^2}}{k_0}h_{18}. 
\label{eq:34a}
\ea 
\end{enumerate}
\subsection{Tensor polarized deuteron target}
The component  of the hadronic tensor, $H_{ij}(S)$, which depends on the deuteron
tensor polarization has the following general structure:
\ba
H_{ij}(S)&=&S_{ab}m_am_b(h_{19}m_im_j+h_{20}n_in_j+
h_{21}\hat k_i\hat k_j+h_{22}\{m,\hat k\}_{ij}+
ih_{23}[m,\hat k]_{ij})+\nn\\
&&
+S_{ab}n_an_b(h_{24}m_im_j+h_{25}n_in_j+h_{26}\hat k_i\hat k_j+
h_{27}\{m,\hat k\}_{ij}+ih_{28}[m,\hat k]_{ij})+ \nn\\
&&
+S_{ab}m_a\hat k_b(h_{29}m_im_j+h_{30}n_in_j+h_{31}\hat k_i\hat k_j+
h_{32}\{m,\hat k\}_{ij}+ih_{33}[m,\hat k]_{ij})+ \nn\\
&&
+S_{ab}m_an_b(h_{34}\{m,n\}_{ij}+h_{35}\{\hat k,n\}_{ij}+
ih_{36}[m,n]_{ij}+ih_{37}[\hat k,n]_{ij})+ \nn\\
&&
+S_{ab}\hat k_an_b(h_{38}\{m,n\}_{ij}+h_{39}\{\hat k,n\}_{ij}+
ih_{40}[m,n]_{ij}+ih_{41}[\hat k,n]_{ij}). 
\label{eq:eq12}
\ea
In this case, the dependence of the polarization observables on the deuteron
tensor polarization is determined by 23 structure functions. The
expressions for these structure functions in terms of the reaction
amplitudes $g_i (i=1-13)$ are given in  Appendix I.

From this expression one can conclude that:

\begin{enumerate}
\item
If the deuteron is tensor polarized so that only $S_{zz}, \ S_{yy}$ and
$(S_{xz}+S_{zx})$ components of the quadrupole polarization tensor are
nonzero, then the dependence of the differential cross section of the
$e^-+d\to e^-+P+d$ reaction on the parameter $\varepsilon$ and
on the azimuthal angle $\varphi$ must be the same as in the case of the
unpolarized target (more exactly, with similar $\varepsilon$-- and $\varphi$--
dependent terms).
\item
 If the deuteron is polarized so that only the $(S_{xy}+S_{yx})$ and
$(S_{yz}+S_{zy})$ components of the quadrupole polarization tensor are nonzero,
then: 

- for $P$-meson production with unpolarized electron beam the typical terms follow $\sin\varphi$ and  $\sin(2\varphi )$ dependencies;

-  for $P$-meson production with longitudinally polarized electron beam the terms do not depend on $\varepsilon$, $\varphi$, and $\cos\varphi$.
\end{enumerate}

In polarization experiments it is possible to prepare the deuteron target
with definite spin projection on some quantization axis. The corresponding
asymmetry is usually defined as
\be
A=\frac{d\sigma (\lambda _d=+1)-d\sigma (\lambda _d=-1)}
{d\sigma (\lambda _d=+1)+d\sigma (\lambda _d=-1)}, 
\label{eq:21}
\ee
where $d\sigma (\lambda _d)$ is the differential cross section of the
$e^-+d\to e^-+P+d$ reaction when the quantization axis for the
deuteron spin (in the $Pd$--pair CMS) coincides with its momentum, i.e.,
the deuteron has helicity $\lambda _d$. From an experimental point of view,
the measurement of an asymmetry is more convenient than a measurement of a
cross section, as most of systematic experimental errors and other
multiplicative factors cancel in the ratio.

The general form of the hadron tensor $H_{ij}(\lambda _d)$, which determines
the differential cross section of the process under consideration for the
case of the deuteron with helicity $\lambda _d$, can be written as
\ba
H_{ij}(\lambda _d&=&\pm 1)=\delta_1\hat k_i\hat k_j+\delta_2m_im_j
+ \delta_3n_in_j+\delta_4\{\hat k,m\}_{ij}+i\delta_5[\hat k,m]_{ij}\pm
 \nn\\
&&
\pm \delta_6\{\hat k,n\}_{ij}\pm i\delta_7[\hat k,n]_{ij}\pm
\delta_8\{m,n\}_{ij}\pm i\delta_9[m,n]_{ij}. 
\label{eq:21a}
\ea
The reaction amplitude is real in the Born (impulse) approximation. So,
assuming the T-invariance of the hadron electromagnetic interactions, we can
do the following statements, according to the deuteron polarization state:

\underline {The deuteron is unpolarized}. Since, in this case,  the hadronic tensor
$H_{ij}(0)$ has to be symmetric (over the $i,j$ indices), the
asymmetry in the scattering of longitudinally polarized electrons vanishes.

\underline {The deuteron is vector polarized}. Since, in this case, the hadronic tensor
$H_{ij}(\xi )$ has to be antisymmetric, then the deuteron vector
polarization can manifest itself in the scattering of longitudinally polarized
electrons. The perpendicular target polarization (normal to the
$\gamma^*+d\rightarrow d+\pi^0$ reaction plane) leads to a correlation of
the following type: $\pm i\lambda\sqrt{2\varepsilon (1-\varepsilon )}
\sin\varphi .$  The longitudinal and transverse (along or perpendicular to the
virtual--photon momentum) target polarization (lying in the
$\gamma^*+d\rightarrow P+d$ reaction plane) leads to two correlations
of the following type : $\mp i\lambda\sqrt{1-\varepsilon ^2}$ and
$\mp i\lambda\sqrt{2\varepsilon (1-\varepsilon )}\cos\varphi .$

\underline {The deuteron is tensor polarized}. The hadronic tensor
$H_{ij}(S)$ is symmetric in this case. In the scattering of longitudinally
polarized electrons the contribution proportional to $\lambda S_{ab}$
vanishes. If the target is polarized so that only the $(S_{xy}+S_{yx})$ or
$(S_{yz}+S_{zy})$ components of the quadrupole polarization tensor are nonzero,
then in the differential cross section only the following two terms are present:
$\varepsilon \sin(2\varphi)$ and  $\sqrt{2\varepsilon (1+\varepsilon )}
\sin\varphi .$ For all other target polarizations the following structures are
present: a term which does not depend on $\varepsilon $ and $\varphi $
variables as well as terms with the following dependencies: $2\varepsilon $,
$\varepsilon cos(2\varphi ) $, and $\sqrt{2\varepsilon (1+\varepsilon )}
cos\varphi .$

The differential cross section of the $P$--meson  production in the
scattering of longitudinally polarized electrons by a tensor polarized deuteron
target (in a coincidence experimental setup) has the following general
structure
\ba
\frac{d^3\sigma}{dE'd\Omega_ed\Omega_{P}} &=&
N\biggl \{\sigma_T+A_{xz}^TQ_{xz}+A_{xx}^T(Q_{xx}-Q_{yy})+A_{zz}^TQ_{zz}+
\nn\\
&&
+\varepsilon \biggl [\sigma_L+A_{xz}^LQ_{xz}+A_{xx}^L(Q_{xx}-Q_{yy})+
A_{zz}^LQ_{zz}\biggr ]+ \nn\\
&&
+\sqrt{2\varepsilon (1+\varepsilon )}\cos\varphi \biggl [
\sigma_I+A_{xz}^IQ_{xz}+A_{xx}^I(Q_{xx}-Q_{yy})+A_{zz}^IQ_{zz}\biggr ]+ \nn\\
&&
+\sqrt{2\varepsilon (1+\varepsilon )}\sin\varphi (
A_{xy}^IQ_{xy}+A_{yz}^IQ_{yz}))+ \nn\\
&&
+\varepsilon \sin(2\varphi )(A_{xy}^PQ_{xy}+A_{yz}^PQ_{yz})+ \nn\\
&&
+\varepsilon \cos(2\varphi )
\biggl [\sigma_P+A_{xz}^PQ_{xz}+A_{xx}^P(Q_{xx}-Q_{yy})+
A_{zz}^PQ_{zz}\biggr ]+\nn\\
&&
+\lambda \sqrt{2\varepsilon (1-\varepsilon )}\sin\varphi \biggl [
\sigma_I'+\bar A_{xz}^IQ_{xz}+\bar A_{xx}^I(Q_{xx}-Q_{yy})+
\bar A_{zz}^IQ_{zz}\biggr ]+ \nn\\
&&+\lambda \sqrt{2\varepsilon (1-\varepsilon )}\cos\varphi \biggl [
\bar A_{xy}^IQ_{xy}+\bar A_{yz}^IQ_{yz}\biggr ]+
\nn\\
&&
+
\lambda \sqrt{1-\varepsilon^2}\cos\varphi \biggl [
A_{xy}^TQ_{xy}+A_{yz}^TQ_{yz}\biggr ]\biggr \}, 
\label{eq:21b}
\ea
where the quantities $Q_{ij} (i,j=x,y,z)$ are the components of the quadrupole
polarization tensor of the deuteron in its rest system (the coordinate system
is specified similarly to the case of the $\pi^0d$--pair CMS). These
components satisfy to the following conditions: $Q_{ij}=Q_{ji}$, $Q_{ii}=0$.
In the derivation of this formula we take into account that $Q_{xx}+Q_{yy}+Q_{zz}=0$.

Thus, in the general case the exclusive cross section of the $P$--meson
production in the scattering of longitudinally polarized electrons by a tensor
polarized deuteron target is determined by 23 independent asymmetries (16(7)
ones in the scattering of unpolarized(longitudinally polarized) electrons)
$A_{ij}^m(W,k^2,\vartheta )$, where $i,j=x,y,z; m=T,P,L,I$. These asymmetries
can be related to the structure functions $h_i$ which are the bilinear
combinations of the 13 independent scalar amplitudes describing the
$\gamma ^* +d\to P+d$ reaction. These relations are:
\ba
A_{xz}^T&=&\gamma_1(h_{29}+h_{30}), \ \
A_{xx}^T=\frac{1}{2}(h_{19}+h_{20}-h_{21}-h_{25}), \ \
A_{zz}^T=-\frac{1}{2}(h_{19}+h_{20}+h_{21}+h_{25}),\nn\\
A_{xz}^L&=&-2\gamma_1\frac{k^2}{k_0^2}h_{31}, \ \
A_{xx}^L=-\frac{k^2}{k_0^2}(h_{21}-h_{26}), \ \
A_{zz}^L=\frac{k^2}{k_0^2}(h_{21}+h_{26}),
\nn\\
A_{xz}^I&=&-2\gamma_1\frac{\sqrt{-k^2}}{k_0}h_{32}, \ \
A_{xx}^I=-\frac{\sqrt{-k^2}}{k_0}(h_{22}-h_{27}),  \ \
A_{zz}^I=\frac{\sqrt{-k^2}}{k_0}(h_{22}+h_{27}), 
\label{eq:AA}\\
A_{xy}^I&=&-2\frac{\sqrt{-k^2}}{k_0}h_{35}, \ \
A_{yz}^I=-2\gamma_1\frac{\sqrt{-k^2}}{k_0}h_{39}, \ \
A_{xy}^P=2h_{34}, \ \ A_{yz}^P=2\gamma_1h_{38},
\nn\\
A_{xz}^P&=&\gamma_1(h_{29}-h_{30}), \ \
A_{xx}^P=\frac{1}{2}(h_{19}-h_{20}-h_{24}+h_{25}), \ \
A_{zz}^P=-\frac{1}{2}(h_{19}-h_{20}+h_{24}-h_{25}),
\nn\\
\bar A_{xz}^I&=&-2\gamma_1\frac{\sqrt{-k^2}}{k_0}h_{33}, \ \
\bar A_{xx}^I=-\frac{\sqrt{-k^2}}{k_0}(h_{23}-h_{28}),  \ \ \bar
A_{zz}^I=\frac{\sqrt{-k^2}}{k_0}(h_{23}+h_{28}),\nn\\
\bar A_{xy}^I&=&-2\frac{\sqrt{-k^2}}{k_0}h_{37}, \ \
\bar A_{yz}^I=-2\gamma_1\frac{\sqrt{-k^2}}{k_0}h_{41}, \ \
A_{xy}^T=2h_{36}, \ \ A_{yz}^T=2\gamma_1h_{40}.\nnñ
\label{eq:AAa}
\ea 
One can see from this formula that the scattering of unpolarized
electrons by a tensor polarized deuteron target with components
$Q_{xy}=Q_{yz}=0$, is characterized by the same $\varphi $-- and
$\varepsilon $--dependences as in the case of the scattering of
unpolarized electrons by the unpolarized deuteron target. If
$Q_{xy}\ne 0, Q_{yz}\ne 0$, then new terms of the type
$\sqrt{2\varepsilon (1+\varepsilon )}\sin\varphi $ and $\varepsilon
\sin(2\varphi )$ are present in the cross section. The asymmetries
with upper indices $T, P (L)$ are determined only by the transverse
(longitudinal) components of the electromagnetic current for the
$\gamma ^* +d\to P+d$ reaction, while the asymmetries with
upper index $I$ are determined by the interference of the
longitudinal and transverse components of the electromagnetic
current.

Using the explicit form for the amplitude of the reaction under consideration
it is easy to obtain the expression for the hadronic tensor $H_{ij}$
in terms of the scalar amplitudes $g_i \ (i=1, ..., 13).$ Appendix I 
contains the formulae for the structure functions $h _i$ in terms of the
scalar amplitudes, which describe the polarization effects in the
$e^-+d\to e^-+P+d$ reaction caused by the deuteron polarization.

Let us stress again that the results listed above have a general nature and
are not related to a particular reaction mechanism. They are valid for the
one--photon--exchange mechanism assuming P-invariance of the hadron
electromagnetic interaction. Their general nature is due to the fact that the 
derivation of these formulae requires only the hadron electromagnetic current
conservation and the fact that the photon has spin one.
\section{Polarization state of scattered deuteron}
Let us consider the general structure of the polarization effects
related to the polarization of the scattered deuteron. The scattered
deuteron spin density matrix can be written as
\be
\label{densitymatrix}
\rho^s_{\mu\nu}=-(g_{\mu\nu}-\frac{p_{2\mu}p_{2\nu}}{M^2})+
\frac{i}{2M}\varepsilon
_{\mu\nu\alpha\beta}\tilde{s}_{\alpha}p_{2\beta}+\tilde{S}_{\mu\nu},
\ee
where $\tilde{s}_{\alpha}$ is the four--vector describing the vector
polarization of the scattered deuteron, $\tilde{s}^2=-1,$
$\tilde{s}\cdot p_2=0$ and $\tilde{S}_{\mu\nu}$ is the tensor
describing the tensor (quadrupole) polarization of the scattered
deuteron, $\tilde{S}_{\mu\nu}=\tilde{S}_{\nu\mu},$
$p_{2\mu}\tilde{S}_{\mu\nu}=0, $ $\tilde{S}_{\mu\mu}=0 $ (due to
these properties the tensor $\tilde{S}_{\mu\nu}$ has only five
independent components). 

In the scattered deuteron rest system all
time components of the tensor $\tilde{S}_{\mu\nu}$ are zero and the
tensor polarization of the scattered deuteron is described by five
independent space components 
$(\tilde{S}_{ij}= \tilde{S}_{ji},\tilde{S}_{ii}=0, i,j=x,y,z).$ 
The four--vector $\tilde{s}_{\alpha}$
is related to the unit vector ${\vec \zeta}$ of the scattered
deuteron vector polarization in its rest system by 
$\tilde{s}_0=-{\vec
q}{\vec \zeta}/M, \ $ ${\vec \tilde{s}}={\vec \zeta}+ {\vec q}{\vec
q}\cdot{\vec \zeta}/M(M+E_2),$ $E_2$ is the scattered deuteron energy in the $\gamma ^*+d\to d+P$ reaction CMS.

The hadronic tensor $H_{ij} (i,j=x,y,z)$ has a linear dependence on the
scattered deuteron polarization parameters and it can be represented
as follows:
\be\label{eq:22}
H_{ij}=H_{ij}(0)+H_{ij}(\zeta)+H_{ij}(\tilde{S}),
\ee
where the term $H_{ij}(0)$ corresponds to the case of the
unpolarized deuteron target and scattered deuteron, and the term
$H_{ij}(\zeta) (H_{ij}(\tilde{S}))$ corresponds to the case of the
vector (tensor)-polarization of the scattered deuteron provided
that target is unpolarized.
\begin{itemize}
\item 
\underline{The scattered deuteron is unpolarized}. The structure of the
tensor $H_{ij}(0)$ is given by Eq. (\ref{eq:8}) with the same structure
functions $h_i, i=1-5$.
\item
\underline{The scattered deuteron is vector polarized}. The structure of
the tensor $H_{ij}(\zeta)$ is given by Eq. (\ref{eq:vecpol}) where it is
necessary to do the following change
$\vec{\xi}\to\vec{\zeta}$ and the structure functions must
be also changed $h_i\to \bar{h}_i, i=6-18$. Therefore, the
dependence of the polarization observables on the vector
polarization of the scattered deuteron is also determined by 13
structure functions. The expressions for the structure functions
$\bar{h}_i$ in terms of the reaction amplitudes $g_i (i=1-13)$ are
given in Appendix I.
The differential cross section of the reaction $d({\vec e},e'P){\vec
d}$, where the electron beam is longitudinally polarized and the
scattered deuteron has vector polarization, can be written as
follows:
\be\label{eq:23}
\frac{d^3\sigma}{dE'd\Omega_ed\Omega_{P}} = \sigma_0\left
[1+\lambda \Sigma_{e}+(P_x^d+\lambda T_x^{ed})\zeta_x+
(P_y^d+\lambda T_y^{ed})\zeta_y+(P_z^d+\lambda T_z^{ed})\zeta_z
\right],
\ee
where $P_i^d (i=x,y,z)$ are the components of the vector
polarization of the scattered deuteron, and $T_i^{ed} (i=x,y,z)$ are
the coefficients of the polarization transfer from the longitudinal
polarization of the electron beam to the vector polarization of the scattered
deuteron. The quantities $\Sigma_{e}$ and $P_i^d$ are T--odd
observables and they are completely determined by the reaction
mechanism beyond the impulse approximation, for example, by the
final--state interaction effects. On the contrary, the quantities
$T_i^{ed}$ are T--even observables and they do not vanish in 
absence of  final--state interaction effects.

The expressions of the $P_i^{d}$ and $T_i^{ed}$ polarization
observables can be also explicitly written as functions of the
azimuthal angle $\varphi $, of the virtual photon linear
polarization $\varepsilon $, and of the contributions of the
longitudinal (L) and transverse (T) components (relative to the
virtual photon momentum ${\vec k}$) of the hadron electromagnetic
current of the $\gamma^*+d\to d+P$ reaction. These expressions can
be obtained from Eqs. (\ref{eq:34},\ref{eq:34a}) with the following substitutions:
$A_i^{d}\to P_i^{d}$, $A_i^{ed}\to T_i^{ed}$,
$A_i^{(IJ)}\to P_i^{(IJ)}$ and $B_i^{(IJ)}\to
C_i^{(IJ)}$, where $I,J=L,T$.

The individual contributions to the components of the vector
polarization and polarization transfer coefficients in terms of the
structure functions $\bar{h_i}$ are given by Eq. (\ref{eq:34a}) where it is
necessary to change $h_i\to \bar{h_i}$ .

At this stage, the general model--independent analysis of the
polarization observables in the reaction $d({\vec e}, e'P)\vec{d}$,
for the case of the vector-polarized scattered deuteron, is
completed. To proceed further in the calculation of the observables,
one needs a model for the reaction mechanism and for the deuteron
structure.

\item 
\underline{The scattered deuteron is tensor polarized}. The general
structure of the tensor $H_{ij}(\tilde{S})$ is the same as given by
Eq. (\ref{eq:eq12}) where it is necessary to do the following change
$S_{ab}\to\tilde{S}_{ab}$ and $h_i\to \bar{h}_i, i=19-41$ for the structure functions. Therefore, the
dependence of the polarization observables on the tensor
polarization of the scattered deuteron is also determined by 23
structure functions. The expressions for the structure functions
$\bar{h}_i$ in terms of the reaction amplitudes $g_i (i=1-13)$ are
given in Appendix I.

The differential cross section of the $P$--meson  production in the
scattering of longitudinally polarized electrons by an unpolarized
deuteron target, when the tensor polarization of the scattered
deuteron is measured (in a coincidence experimental setup), has
the following general structure:
\ba
\frac{d^3\sigma}{dE' d\Omega_e d\Omega_{P}} &=&
N\biggl \{\sigma_T+P_{xz}^T\tilde{Q}_{xz}+P_{xx}^T(\tilde{Q}_{xx}-\tilde{Q}_{yy})+P_{zz}^T\tilde{Q}_{zz}+
\nn\\
&&
+\varepsilon \biggl [\sigma_L+P_{xz}^L\tilde{Q}_{xz}+P_{xx}^L(\tilde{Q}_{xx}-\tilde{Q}_{yy})+
P_{zz}^L\tilde{Q}_{zz}\biggr ]+ \nn\\
&&
+\sqrt{2\varepsilon (1+\varepsilon )}\cos\varphi \biggl [
\sigma_I+P_{xz}^I\tilde{Q}_{xz}+P_{xx}^I(\tilde{Q}_{xx}-\tilde{Q}_{yy})+P_{zz}^I\tilde{Q}_{zz}\biggr ]+ \nn\\
&&
+\sqrt{2\varepsilon (1+\varepsilon )}\sin\varphi (
P_{xy}^I\tilde{Q}_{xy}+P_{yz}^I\tilde{Q}_{yz})+
\varepsilon \sin(2\varphi )(P_{xy}^P\tilde{Q}_{xy}+P_{yz}^P\tilde{Q}_{yz})+ \nn\\
&&
+\varepsilon \cos(2\varphi )
\biggl [\sigma_P+P_{xz}^P\tilde{Q}_{xz}+P_{xx}^P(\tilde{Q}_{xx}-\tilde{Q}_{yy})+
P_{zz}^P\tilde{Q}_{zz}\biggr ]+\nn\\
&&
+\lambda \sqrt{2\varepsilon (1-\varepsilon )}\sin\varphi \biggl [
\sigma_I'+\bar P_{xz}^I\tilde{Q}_{xz}+\bar P_{xx}^I(\tilde{Q}_{xx}-\tilde{Q}_{yy})+
\bar P_{zz}^I\tilde{Q}_{zz}\biggr ]+ \nn\\
&&+\lambda \sqrt{2\varepsilon (1-\varepsilon )}\cos\varphi \biggl [
\bar P_{xy}^I\tilde{Q}_{xy}+\bar P_{yz}^I\tilde{Q}_{yz}\biggr ]+
\nn\\
&&
+
\lambda \sqrt{1-\varepsilon^2}\cos\varphi \biggl [
P_{xy}^T\tilde{Q}_{xy}+P_{yz}^T\tilde{Q}_{yz}\biggr ]\biggr \}, 
\label{eq:24a}
\ea
where the quantities $\tilde{Q}_{ij} (i,j=x,y,z)$ are the components
of the quadrupole polarization tensor of the scattered deuteron in
the $Pd$--pair CMS. These components satisfy to the following
conditions: 
\be
\tilde{Q}_{ij}=\tilde{Q}_{ji}, \ 
d\tilde{Q}_{zz}+u\tilde{Q}_{xx}+\gamma^2_2\tilde{Q}_{yy}-
2z(1+\gamma_2)\tilde{Q}_{xz}=0,
\label{eq:cond}
\ee 
where
$d=cos^2\vartheta+sin^2\vartheta \gamma^2_2 , \
u=sin^2\vartheta+cos^2\vartheta \gamma^2_2,$
$z=(\gamma_2-1)cos\vartheta \sin\vartheta , \gamma_2=E_2/M$. Eq. (\ref{eq:24a}) takes into account the last condition.
\end{itemize}

Thus, in the general case the exclusive cross section of the
$P$--meson production in the scattering of longitudinally polarized
electrons by unpolarized deuteron target, when the tensor
polarization of the scattered deuteron is measured (in the
coincidence experimental setup), is determined by 23 independent
functions (16 (7) ones in the scattering of unpolarized
(longitudinally polarized) electrons) $P_{ij}^m(W,k^2,\vartheta )$,
where $i,j=x,y,z; m=T,P,L,I$. These asymmetries can be related to the
structure functions $\bar{h}_i$ which are the bilinear combinations
of the 13 independent scalar amplitudes describing the $\gamma ^*
+d\to P+d$ reaction. These relations are:
\ba
rP_{xz}^T&=&r(\bar{h}_{29}+\bar{h}_{30})-2z(1+\gamma_2)(\bar{h}_{19}+\bar{h}_{20}+
\bar{h}_{24}+\bar{h}_{25}), \nn\\
rP_{xx}^T&=&u(\bar{h}_{24}+\bar{h}_{25})-\gamma^2_2(\bar{h}_{19}+\bar{h}_{20}),
\ rP_{zz}^T=d(\bar{h}_{19}+\bar{h}_{20}+
\bar{h}_{24}+\bar{h}_{25}), \nn\\
rP_{xz}^L&=&2\frac{Q^2}{k_0^2}[r\bar{h}_{31}-2z(1+\gamma_2)(\bar{h}_{21}+\bar{h}_{26})],\ 
rP_{xx}^L=2\frac{Q^2}{k_0^2}(u\bar{h}_{26}-\gamma^2_2\bar{h}_{21}),
\nn\\
rP_{zz}^L&=&2\frac{Q^2}{k_0^2}d(\bar{h}_{21}+\bar{h}_{26}), \ \
rP_{xz}^I=-2\frac{\sqrt{Q^2}}{k_0}[r\bar{h}_{32}-
2z(1+\gamma_2)(\bar{h}_{22}+\bar{h}_{27})], 
\nn\\ 
rP_{xx}^I&=&-2\frac{\sqrt{Q^2}}{k_0}(u\bar{h}_{27}-\gamma^2_2\bar{h}_{22}),
rP_{zz}^I=-2\frac{\sqrt{Q^2}}{k_0}d(\bar{h}_{22}+\bar{h}_{27}), \ \
\label{eq:25}\\
P_{xy}^I&=&-2\frac{\sqrt{Q^2}}{k_0}\bar{h}_{35}, \ \
P_{yz}^I=-2\frac{\sqrt{Q^2}}{k_0}\bar{h}_{39}, \ \
P_{xy}^P=2\bar{h}_{34},\ P_{yz}^P=2\bar{h}_{38}, \nn\\
rP_{xz}^P&=&r(\bar{h}_{29}-\bar{h}_{30})
-2z(1+\gamma_2)(\bar{h}_{19}-\bar{h}_{20}+
\bar{h}_{24}-\bar{h}_{25}), \ \
 \nn\\
rP_{xx}^P&=&u(\bar{h}_{24}-\bar{h}_{25})-\gamma^2_2(\bar{h}_{19}-\bar{h}_{20}),
\nn\\
rP_{zz}^P&=&d(\bar{h}_{19}-\bar{h}_{20}+\bar{h}_{24}-\bar{h}_{25}), \
\ r\bar P_{xz}^I=-2\frac{\sqrt{Q^2}}{k_0}[r\bar{h}_{33}-
2z(1+\gamma_2)(\bar{h}_{23}+\bar{h}_{28})],
\nn\\
r\bar
P_{xx}^I&=&-2\frac{\sqrt{Q^2}}{k_0}(u\bar{h}_{28}-\gamma^2_2\bar{h}_{23}),
\ \ r\bar
P_{zz}^I=-2\frac{\sqrt{Q^2}}{k_0}d(\bar{h}_{23}+\bar{h}_{28}), \ \
\nn\\
\bar P_{xy}^I&=&-2\frac{\sqrt{Q^2}}{k_0}\bar{h}_{37},
\bar P_{yz}^I=-2\frac{\sqrt{Q^2}}{k_0}\bar{h}_{41}, \ \
P_{xy}^T=2\bar{h}_{36}, \ \ P_{yz}^T=2\bar{h}_{40},\nn
\ea
where $r=-(u+\gamma^2_2)$.

The results listed above have a general nature and are not related
to a particular reaction mechanism. They are valid for the
one--photon--exchange mechanism assuming P-invariance of the hadron
electromagnetic interaction. Their general nature is due to the fact
that derivation of these results requires only the hadron
electromagnetic current conservation and the fact that the photon
has spin one.
\section{Coherent production of pseudoscalar meson in deuteron
photodisintegration process}
Let us consider the particular case of the coherent photoproduction
of the pseudoscalar meson on the deuteron target
\be\label{eq:21c}
\gamma (k)+d(p_1)\to P(q)+d(p_2),
\ee
where the four--momenta of the particles are given in the brackets. Of course,
all observables for this reaction can be obtained using the formulae presented
above for the case of the virtual photon, but it is rather tedious procedure.
So, it is worth to have the expressions for the differential cross section and
various polarization observables which are suitable for the analysis of the
data on this reaction.

The matrix element of this reaction can be written as
\be\label{eq:22a}
M=eA_{\mu}J_{\mu}=-eA_{i}J_{i},
\ee
where $A_{\mu}$ is the photon polarization four--vector and we use the
transverse gauge: ${\vec k}\cdot{\vec A}=0$ (${\vec k}$ is the photon momentum),

The differential cross section in CMS (not averaged over the spins of the
initial particles) can be written as
\be
\label{eq:23a}
\frac{d\sigma}{d\Omega}=\frac{\alpha}{8\pi}\frac{q}{W}
\frac{1}{W^2-M^2}\rho_{ij}H_{ij},
\ee
where $\rho_{ij}=A_iA_j^*$ and hadronic tensor is determined as
$H_{ij}= J_iJ_j^*$. The quantities which are not redefined in this
section have the same meaning as in the previous sections.

In the reaction CMS, the quantity $J_i$ can be represented as
\be
J_{i}=m_iA+n_iB,
\label{eq:24}
\ee
where
\ba
 A&=&g_1{\vec m}\cdot {\vec
U}_1{\vec n}\cdot {\vec U}_2^*+ g_2\hat {\vec k}\cdot {\vec
U}_1{\vec n}\cdot {\vec U}_2^*+ g_3{\vec n}\cdot {\vec U}_1{\vec
m}\cdot {\vec U}_2^*+ g_4{\vec n}\cdot {\vec U}_1\hat {\vec k}\cdot
{\vec U}_2^*, \nn
\\
B&=&g_5{\vec m}\cdot {\vec U}_1{\vec m}\cdot {\vec U}_2^*+ g_6{\vec
n}\cdot {\vec U}_1{\vec n}\cdot {\vec U}_2^*+ g_7\hat {\vec k}\cdot
{\vec U}_1\hat {\vec k}\cdot {\vec U}_2^*+ g_8{\vec m}\cdot {\vec
U}_1\hat {\vec k}\cdot {\vec U}_2^*+g_9\hat {\vec k}\cdot {\vec
U}_1{\vec m}\cdot {\vec U}_2^*.\nn
\ea
In this case the nine scalar amplitudes depend on two variables $W$
and $\vartheta $ (energy and scattering angle) instead of three ones
for the case of the pseudoscalar meson electroproduction ($k^2\ne
0$).

The hadronic tensor $H_{ij} (i,j=x,y,z)$ can be also represented in the form
given by Eq. (\ref{eq:hadrontensor}) where each term corresponds to the definite polarization
state of the deuteron target, provided that the scattered deuteron is unpolarized.

Let us consider the polarization observables of the $\gamma +d\to P+d$ reaction 
which correspond to each contribution of the hadronic tensor $H_{ij}$.

\begin{itemize}
\item
\underline {The deuteron target is unpolarized}. The general structure of the
hadronic tensor for the case of unpolarized deuteron target has the following form
\be\label{eq:25a}
H_{ij}(0)=h_1m_im_j+h_2n_in_j,
\ee
where the structure functions $h_i$, for the case of the
photoproduction of the P-meson on the deuteron, can be expressed in
terms of the $\gamma +d\to P+d$ reaction scalar amplitudes $g_i
(i=1-9)$ using the expressions in Appendix I, where it is
necessary to cancel the four amplitudes $g_i (i=10-13)$, since they
correspond to the absorption of a virtual photon with longitudinal
polarization.

Then the differential cross section of the $\gamma +d\to P+d$ reaction for
the case of unpolarized particles can be written as:
\be
\label{eq:26}
\frac{d\sigma_{un}}{d\Omega}=N(h_1+h_2), \ \
N=\frac{\alpha}{16\pi}\frac{q}{W}\frac{1}{W^2-M^2}.
\ee
\item 
Let us consider the case when the photon beam is polarized. The general
expression of the photon polarization vector is determined by two
real parameters $\beta $ and $\delta $ and it can be written as
\cite{Akhiezer:1965}
\be\label{eq:27}
{\vec A}=\cos\beta {\vec m}+\sin\beta \exp(i\delta ){\vec n}.
\ee
If the parameter $\delta$ vanishes,  $\delta =0$, then this photon polarization vector describes the
linear polarization state of the photon at an angle $\beta $ with respect to the $x$ axis.
The parameters $\beta =\pi /4$ and $\delta =\pm \pi /2$ denote circular
polarization of the photon. Arbitrary $\beta $ and $\delta $ correspond to the
elliptic polarization of the photons.

The differential cross section when only the  photon beam
is polarized has the following form:
\be\label{eq:28}
\frac{d\sigma}{d\Omega}=\frac{d\sigma_{un}}{d\Omega}
(1+A_{\perp}\cos2\beta ),
\ee
where $A_{\perp}$ is the asymmetry due to the linear polarization
of the photon beam. It is defined as: 
\be\label{eq:29}
A_{\perp}=
\frac{d\sigma /d\Omega (\beta =0^\circ)-d\sigma /d\Omega (\beta =90^\circ)}
{d\sigma /d\Omega (\beta =0^\circ)+d\sigma /d\Omega (\beta =90^\circ)},
\ee
and it has the following form in terms of the structure functions:
\be
\label{eq:30}
\frac{d\sigma_{un}}{d\Omega}A_{\perp}=N(h_1-h_2) \ \ or \ \
A_{\perp}=\frac{h_1-h_2}{h_1+h_2}.
\ee
Note that the circular polarization of the photon beam does not
contribute to the differential cross section due to the
P--invariance of the hadron electromagnetic interaction.
\item 
\underline {The deuteron target is vector polarized}. In the case of the pseudoscalar meson photoproduction, the part of the hadronic
tensor which depends on the deuteron vector polarization is determined by six
structure functions. It can be written as:
\ba
H_{ij}(\xi )&=&{\vec\xi }{\vec m}(h_{6}\{m,n\}_{ij}+ih_{8}[m,n]_{ij})+
{\vec\xi }{\vec m}(h_{10}m_im_j+h_{11}n_in_j)+
\nn\\
&&
{\vec\xi }\hat {\vec k}(h_{15}\{m,n\}_{ij}+ih_{17}[m,n]_{ij}). 
\label{eq:31}
\ea
Therefore, for the $\gamma +d\to P+d$ reaction, the dependence of the polarization observables on the deuteron vector polarization is determined
by six structure functions.

The part of the differential cross section of the $\gamma +d\to P+d$
reaction which depends on the deuteron vector polarization, for the case of
arbitrarily polarized photon, can be written as
\ba
\frac{d\sigma_v}{d\Omega}&=&\frac{d\sigma_{un}}{d\Omega}
\biggl [A_{y}\xi_y+C_y^l\cos2\beta \xi_y+\sin2\beta \cos\delta (C_x^l\xi_x+
C_z^l\xi_z)+\nn\\
&&
+\sin 2\beta \sin\delta (C_x^c\xi_x+C_z^c\xi_z)\biggr ],
\label{eq:32}
\ea
where $A_{y}$ is the asymmetry due to the vector polarization of the
deuteron target, provided that the photon is unpolarized (the so called single
target asymmetry). This asymmetry is due to the normal (to the reaction plane)
component of the polarization vector ${\vec \xi}$ describing the vector
polarization of the target. If the reaction amplitudes are real functions
(as, for example, in the impulse approximation), then this asymmetry is equal
to zero. The quantities $C^l_{x,y,z} (C^c_{x,z})$ are the correlation
coefficients due to the vector polarization of the deuteron target when the 
photon is linear (circularly) polarized. The correlation coefficients
$C^l_{x,y,z}$ are zero when the amplitudes  are real. The correlation coefficients
$C^c_{x,z}$ are determined by the components of the polarization vector lying
in the reaction plane and these coefficients are nonzero, in general, for real amplitudes. 
All these polarization observables can be expressed in
terms of the structure functions $h_i$ and they are
\ba
\frac{d\sigma_{un}}{d\Omega}A_{y}&=&N(h_{10}+h_{11}), \ \
\frac{d\sigma_{un}}{d\Omega}C_{y}^l=N(h_{10}-h_{11}),
\nn\\
\frac{d\sigma_{un}}{d\Omega}C_{x}^l&=&2Nh_{6}, \ \
\frac{d\sigma_{un}}{d\Omega}C_{z}^l=2Nh_{15}, \nn\\
\frac{d\sigma_{un}}{d\Omega}C_{x}^c&=&2Nh_8, \ \
\frac{d\sigma_{un}}{d\Omega}C_{z}^c=2Nh_{17}. 
\label{eq:33}
\ea
\item
\underline {The deuteron target is tensor polarized}.
The part of the hadronic tensor which depends on the tensor (quadrupole)
polarization of the deuteron target is determined by 10 structure functions
for the case of  real photons and its general structure is
\ba
H_{ij}(S)&=&S_{ab}m_am_b(h_{19}m_im_j+h_{20}n_in_j)+
S_{ab}n_an_b(h_{24}m_im_j+h_{25}n_in_j)+
\nn\\
&&+S_{ab}\hat k_am_b(h_{29}m_im_j+h_{30}n_in_j)+
S_{ab}m_an_b(h_{34}\{m,n\}_{ij}+ 
\nn\\
&&
+ih_{36}[m,n]_{ij})+
S_{ab}\hat k_an_b(h_{38}\{m,n\}_{ij}+ih_{40}[m,n]_{ij}). 
\label{eq:34H}
\ea
Thus, for the $\gamma +d\to P+d$ reaction, the dependence of the polarization observables on the deuteron tensor (quadrupole) polarization
is completely determined by 10 structure functions.

For the case of arbitrarily polarized photons, the part of the differential cross section which depends on the deuteron tensor polarization, can be written  as: 
\ba
\frac{d\sigma_t}{d\Omega}&=&\frac{d\sigma_{un}}{d\Omega}
\biggl \{A_{zz}Q_{zz}+A_{xx}(Q_{xx}-Q_{yy})+A_{xz}Q_{xz}+\cos 2\beta
\biggl [C^l_{zz}Q_{zz}+
\nn\\
&&
+C^l_{xx}(Q_{xx}-Q_{yy})+C^l_{xz}Q_{xz}\biggr ]+
\sin 2\beta \cos\delta (C^l_{xy}Q_{xy}+C^l_{yz}Q_{yz})+ \nn\\
&&
+\sin 2\beta \sin\delta (C^c_{xy}Q_{xy}+C^c_{yz}Q_{yz})\biggr \}, 
\label{eq:35}
\ea
where $A_{zz}, A_{xx}$ and $A_{xz}$ are the asymmetries due to the tensor
polarization of the deuteron target when the photon is unpolarized. These
asymmetries are non-zero, in general case, if the reaction amplitudes are 
real functions (as, for example, in the impulse approximation) in contrast to
the $A_y$ asymmetry. The quantities $C^l_{zz}, C^l_{xx}, C^l_{xz}, C^l_{xy}$
and $C^l_{yz}$ are the correlation coefficients due to the tensor polarization
of the deuteron target when the photon is linear polarized (they are also
can be non--zero if the reaction amplitudes are the real functions). The quantities
$C^c_{xy}$ and $C^c_{yz}$ are the correlation coefficients determined
by the tensor polarization of the deuteron target and the circular polarization of
the photon (they originate from reaction mechanisms beyond the
impulse approximation, for example, by the final--state interaction effects).
All these polarization observables can be expressed in terms of the structure
functions $h_i $ and they are:
\ba
\sigma A_{zz}&=&-\frac{N}{2}(h_{19}+h_{20}+h_{24}+h_{25}), \nn\\
\sigma A_{xx}&=&\frac{N}{2}(h_{19}+h_{20}-h_{24}-h_{25}), \nn\\
\sigma A_{xz}&=&N\gamma_1(h_{29}+h_{30}), 
\nn\\
\sigma C^l_{zz}&=&\frac{N}{2}(h_{29}+h_{25}-h_{19}-h_{24}),  \nn\\
\sigma C^l_{xx}&=&\frac{N}{2}(h_{19}+h_{25}-h_{20}-h_{24}),  \nn\\
\sigma C^l_{xz}&=&N\gamma_1(h_{29}-h_{30}),\nn\\
\sigma C^l_{xy}&=&2Nh_{34}, \ \ \sigma C^l_{yz}=2N\gamma_1h_{38}, \nn\\
\sigma C^c_{xy}&=&2Nh_{36}, \ \ \sigma C^c_{yz}=2N\gamma_1h_{40},\label{eq:36}
\ea
with
\be 
\gamma_1=\frac{W^2+M^2}{2MW}, ~
\sigma =\displaystyle\frac{d\sigma_{un}}{d\Omega} =N(h_1+h_2).\nn
\ee
\end{itemize}
\section{Helicity amplitudes}
Sometimes it is more convenient to use the helicity amplitudes formalism. Let
us introduce the set of helicity amplitudes 
$f_{\lambda\lambda '} (k^2, W, \vartheta )$ (where $\lambda$ and $\lambda '$ are the
helicities of the initial ($\gamma ^*+d$) and final ($d+P$) states)
and define the amplitudes 
\be
h_{\lambda\lambda '}=< \lambda _{\Delta}, \lambda _N|T|\lambda _{\gamma},
\lambda _d>={\vec \chi}_{2}^*(\lambda _{\Delta}){\vec F}
(\lambda _{\gamma}, \lambda _d)\chi _1^c(\lambda _N), 
\ee
where $\lambda _{\gamma},
\lambda _d, \lambda' _d $  are the helicities of the virtual
photon, initial and scattered deuteron respectively, with $\lambda
=\lambda _{\gamma}- \lambda _d$ and $\lambda '= \lambda' _{d}$. 
We
choose the following convention:
\ba
 f_1&=&<+|T|++>=\frac{i}{2\sqrt{2}}[g_1+g_6+\cos\vartheta (g_5-g_3)+
\sin\vartheta (g_4-g_8)], \nn\\
f_2&=&<-|T|++>=\frac{i}{2\sqrt{2}}[g_1+g_6-\cos\vartheta (g_5-g_3)-
\sin\vartheta (g_4-g_8)], \nn\\
f_3&=&<0|T|++>=\frac{i}{2}\frac{E_2}{M}[\sin\vartheta (g_5-g_3)+
\cos\vartheta (g_8-g_4)], \nn\\
f_4&=&<+|T|+->=\frac{i}{2\sqrt{2}}[g_6-g_1-\cos\vartheta (g_3+g_5)+
\sin\vartheta (g_4+g_8)], \nn\\
f_5&=&<-|T|+->=\frac{i}{2\sqrt{2}}[g_6-g_1+\cos\vartheta (g_3+g_5)-
\sin\vartheta (g_4+g_8)], \nn\\
f_6&=&<0|T|+->=-\frac{i}{2}\frac{E_2}{M}[\sin\vartheta (g_3+g_5)+
\cos\vartheta (g_4+g_8)], \nn\\
f_7&=&<+|T|+0>=\frac{i}{2}\frac{E_1}{M}[g_2-\sin\vartheta g_7+
\cos\vartheta g_9], \nn\\
f_8&=&<-|T|+0>=\frac{i}{2}\frac{E_1}{M}[g_2+\sin\vartheta g_7-
\cos\vartheta g_9], \nn\\
f_9&=&<0|T|+0>=\frac{i}{\sqrt{2}}\frac{E_1E_2}{M^2}[\cos\vartheta g_7+
\sin\vartheta g_9], \nn\\
f_{10}&=&<+|T|0+>=-\frac{i}{2}\frac{k_0}{\sqrt{Q^2}}[g_{10}-
\cos\vartheta g_{12}+\sin\vartheta g_{13}], \nn\\
f_{11}&=&<-|T|0+>=-\frac{i}{2}\frac{k_0}{\sqrt{Q^2}}[g_{10}+
\cos\vartheta g_{12}-\sin\vartheta g_{13}], \nn\\
f_{12}&=&<0|T|0+>=\frac{i}{\sqrt{2}}\frac{E_2}{M}\frac{k_0}{\sqrt{Q^2}}
[\cos\vartheta g_{13}+\sin\vartheta g_{12}], \nn\\
f_{13}&=&<+|T|00>=-\frac{i}{\sqrt{2}}\frac{E_1}{M}\frac{k_0}{\sqrt{Q^2}}
g_{11}.
 \label{Eq:Hij}
\ea
 where $E_2=(W^2-M_P^2+M^2)/2W$ is the energy of the scattered deuteron in
the reaction CMS.

At this stage, the general model--independent analysis of the polarization observables
for pseudoscalar meson photo--production  is completed. To proceed further in the calculation of the observables, one needs
a model for the reaction mechanism and for the deuteron structure.

\section{Model, kinematics, and results}

In order to illustrate the derived formalism with numerical results it is necessary to calculate the elementary amplitudes in frame of a 
model describing the structure of the involved hadrons. Following Ref.  \cite{Rekalo:2002km} we use the impulse approximation for the deuteron, and consider a model for the interaction of the virtual photon with the nucleon. The neutron and proton structure is parametrized in terms of electromagnetic form factors and deuteron wave functions. The Bonn \cite{PhysRevC.63.024001} or the Paris \cite{Lacombe:1980dr} nucleon-nucleon potentials, were considered in Ref. \cite{Rekalo:2002km} as they give the most different values for the observables. It was shown that other recent potentials based on the Argonne \cite{Wiringa:1984tg} and Reid \cite{Reid:1968sq} potentials give indeed intermediate values. 
The photon interaction is described in frame of an effective Lagrangian model, considering nucleon and $\Delta$-exchange in $s$-channel, nucleon exchange in $t$-channel and $\pi$,$\rho$ and $\omega$-mesons  exchange in $u$-channel. The details of the model are given in Ref. \cite{Rekalo:2002km}. In principle, in the near threshold region, for $\gamma (\gamma^*)+d\to d+\pi^0$,
rescattering effects may play an important role,  for pion S-state 
electroproduction. However, it has been shown in a model independent 
way based only on the Pauli principle, that the main rescattering 
contribution from the two step process: 
$\gamma +d\to p+p+\pi^-(\mbox{and}~ n+n+\pi^+)\to d+\pi^0$ vanishes, when the 
two 
nucleons in the $NN\pi$-intermediate state are on mass shell \cite{Rekalo:2001he}.

As stated in the introduction, the main purpose of this paper is the general and model independent derivation of polarized and unpolarized observables for pion electro and photo-production on the deuteron. Therefore, we give an example of the behavior of some of the observables and do not discuss in extent their dependence on the ingredients of the model.

The experimental detection of the pion and the scattered electron, for a definite beam energy, allows to fully determine 
the kinematics of the $e+d\to e+d +\pi^0$ reaction, i.e., $\gamma^*+d \to d +\pi^0$,  that depends on three kinematical variables. Let us choose 
\begin{itemize}
\item $k^2$: the four momentum squared of the $\gamma^*$, 
\item $s=W^2=(k+p_1)^2=(q+p_2)^2$: the total energy in CMS of the $\gamma^*d$  system,
\item $\vartheta$: the $\pi^0$ emission angle in the CMS of the reaction $\gamma^* + d\to d+\pi^0$.
 \end{itemize}
As shown above, there are 13 independent amplitudes for the reaction $\gamma^* +d\to d+\pi^0$: $g_i(k^2,W,\vartheta),i=1-13$, 
that are functions of these three variables.

In the assumption of impulse approximation, the $\gamma^*$ interacts  on the bare nucleon inside the deuteron either the proton or the neutron, while the other nucleon stays as a spectator. There are six amplitudes for the reaction $\gamma^*+N\to N+\pi^0$, depending also on three variables that can not be connected in a unique way to the previous set. The problem is that one has to make an assumption how the momentum $Q^2$ is transferred to one of the nucleons inside the deuteron, while this nucleon have itself a Fermi momentum.

It seems reasonable to calculate these amplitudes at the same values of the two variables, $k^2$ and $t$. The choice is open for the total energy,
$s=(k+p_1)^2=(k+M)^2 \ne s_1=(k+p_N)^2=(k+m)^2$. The value of the $\pi^0$ emission angle in the CMS of the reaction $\gamma^*+ N\to N+\pi^0$, $\theta_{\pi}$,  calculated from $s_1$ may fall outside the kinematical limits. However, one can increase $s_1$, what can be physically understood taking into account the Fermi-motion of the nucleon in the deuteron (see discussion in Ref. \cite{Rekalo:2002km},  page 11).
 
Therefore  the six independent amplitudes for $\gamma^*+ N \to N+\pi^0$, depend on three kinematical variables: $f^N_i(k^2,W_N, \theta_{\pi}),$ i=1-6, where $s_N=W_N^2$ is the total energy of the $\gamma^*N$  (or $\pi^0 N$).

In Ref. \cite{Rekalo:2002km} the structure functions and the observables have been calculated as a function of these last variables, because it was straightforward to implement the nucleon electro-production model. However, the experimentalists will measure the electron, deuteron and pion in the Lab system. They will have access to $t, W, \cos\vartheta, \cos\theta $ but not to: $ s_N $, and in general to the participant nucleon. 

Therefore we choose to  fix $\vec p_1+\vec k$=0, $s_1=k_0+m$ in a near threshold kinematics, where all phase space is available for the proton: s=4.1 GeV, $0.5\le t \le  2.5 $GeV$^2$ and calculate the structure functions and some polarization observables as a function of the pion angle in CMS of $\gamma^* d $ system, $\cos \vartheta$.

Let us stress that this application is given as an example. The general formalism derived in this paper allows to calculate the observables for any energy and for any kinematics, in frame of the one photon exchange approximation, implementing any suitable model for the deuteron structure and for the reaction. 

\subsection{Numerical results}

		
We choose to illustrate the  results  as a function of $\cos\vartheta$, at the total energy $W$=2.5 GeV for four values of the momentum transfer in the range $0.5 \le -k^{2} \le 2$ GeV$^{2}$. These conditions do not violate the energy and momentum conservation in the $\gamma^{*}d$ and $\gamma^{*}p$ systems in the kinematical range of all the variables considered here. The numerical calculations are done for the Paris potential, for the dipole parametrization of the proton form factor, whereas the electric neutron form factor is set to zero.  
\subsection{Unpolarized Structure Functions}
\begin{figure}
	        \includegraphics [height=10.5cm]{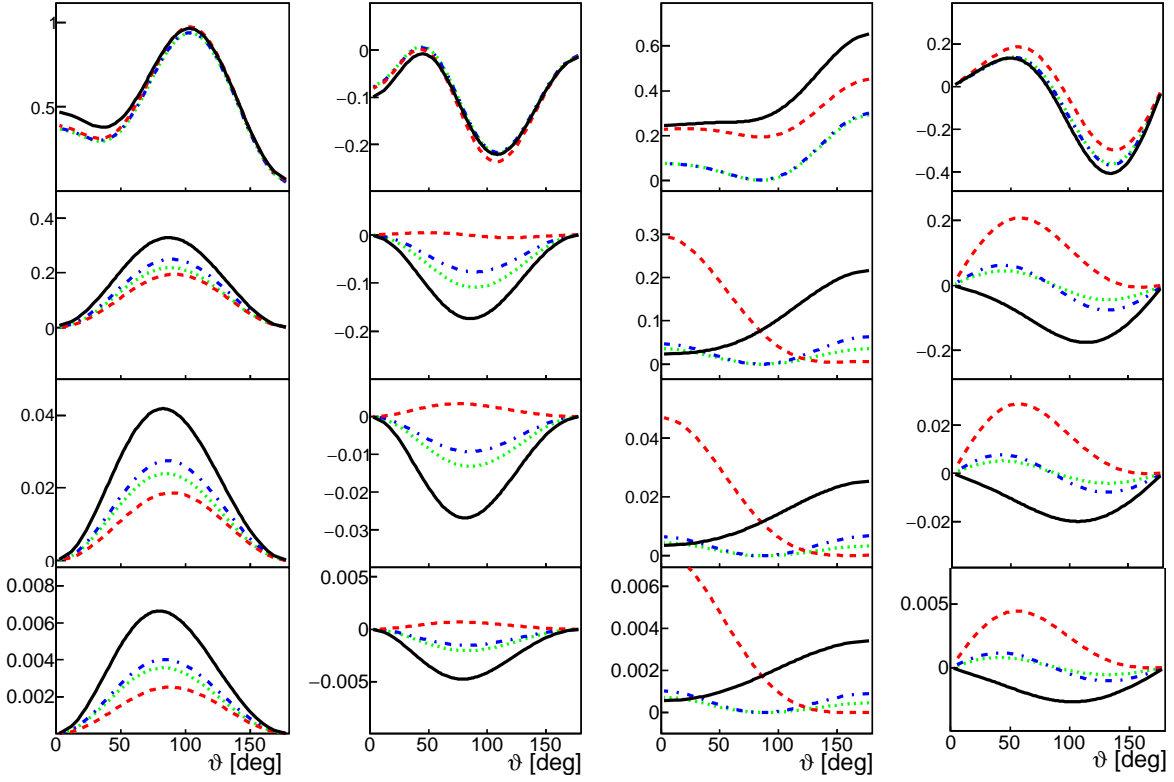}
	        \caption{ $\vartheta$-dependence of the functions  $H_{xx}+H_{yy}$,  $H_{xx}-H_{yy}$, $H_{zz}$ and  $H_{xz}+H_{zx}$ (respectively from left to right) for $W$=2.5 GeV. From top to bottom each plot corresponds to $-k^{2}$= 0.5, 1, 1.5 and 2 GeV$^{2}$. The different lines illustrate the  considered contributions for $\pi^{0}$ production: $\Delta$ (green, dotted line), $\Delta+s+u$ (blue, dash-dotted line), $\Delta+s+u-\omega$ (red, dashed line), $\Delta+s+u+\omega$ (black, solid line).}
	         \label{Fig:HLT} 
	  \end{figure}

The four structure functions that define the unpolarized cross section are shown in Fig. \ref{Fig:HLT}. 
From top to bottom each plot corresponds to $-k^{2}$= 0.5, 1, 1.5 and 2 GeV$^{2}$, $H_{xx}+H_{yy}$,  $H_{xx}-H_{yy}$, $H_{zz}$ and  $H_{xz}+H_{zx}$. The different lines illustrate  the  considered contributions to $\pi^{0}$ production: $\Delta$ (green, dotted line), $\Delta+s+u$ (blue, dash-dotted line), $\Delta+s+u-\omega$ (red, dashed line), $\Delta+s+u+\omega$ (black, solid line). 

\subsection{Observables for a polarized deuteron target}

\begin{figure}
	        \includegraphics [height=10.5cm]{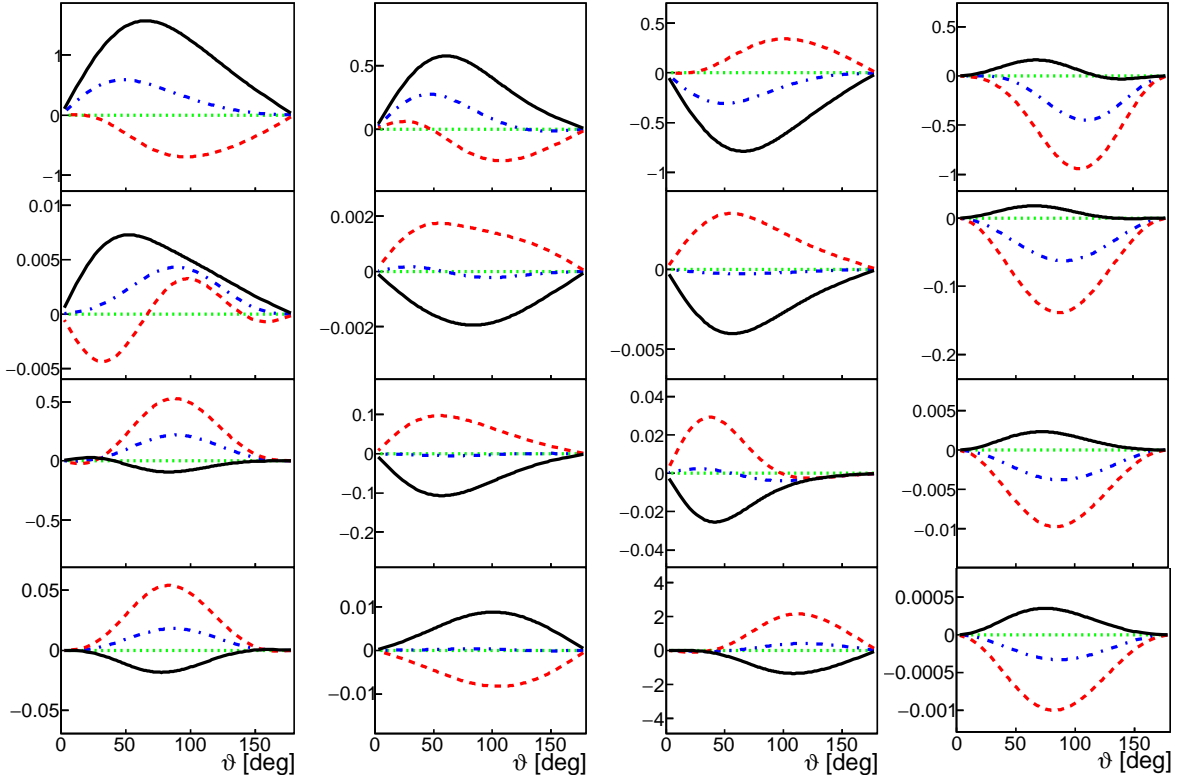}
	        \caption{ $\vartheta$-dependence of the functions  $A_{x}^{(TT)}$, $A_{y}^{(TT)}$, $\bar{A}_{y}^{(TT))}$ and $A_{z}^{(TT)}$  from left to right. Notations as in Fig. \protect\ref{Fig:HLT}.}
	         \label{Fig:ATT} 
	  \end{figure}

\begin{figure}
	        \includegraphics [height=10.5cm]{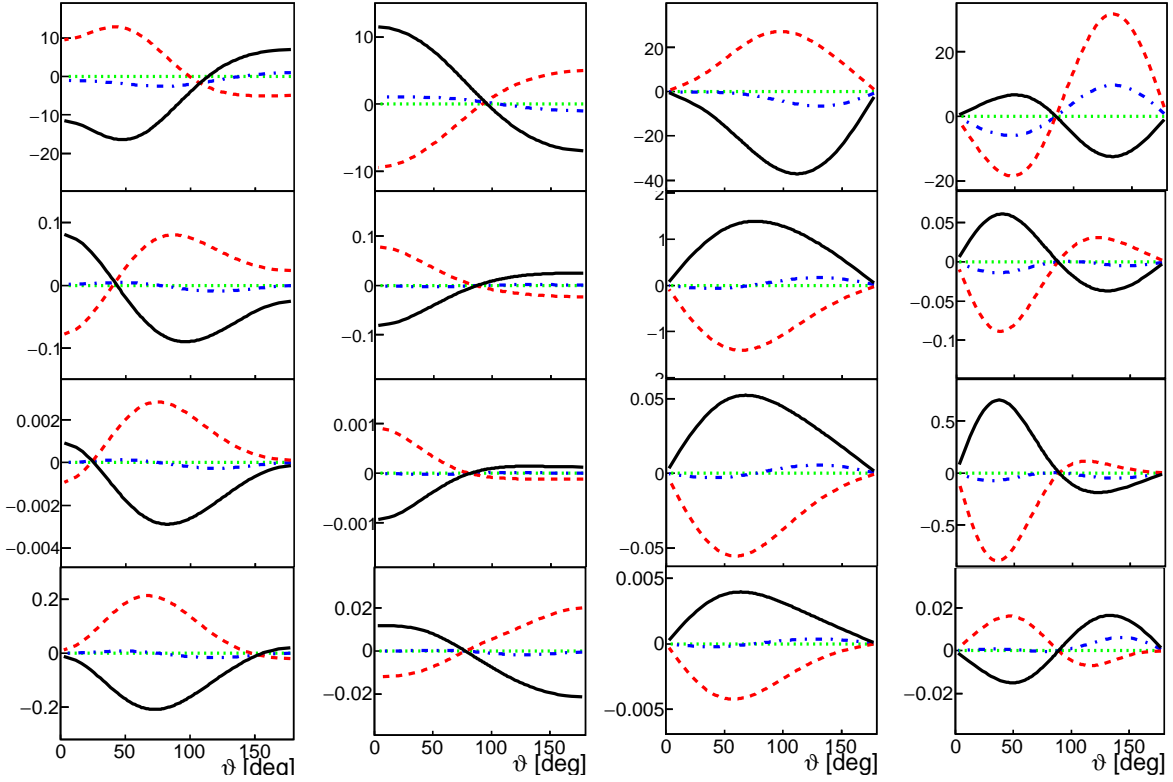}
	        \caption{$\vartheta$-dependence of the functions $A_{x}^{(LT)}$, $A_{y}^{(LT)}$, $A_{z}^{(LT)}$, $A_{y}^{(LL)}$, from left to right. Notations as in Fig. \protect\ref{Fig:HLT}. }
	         \label{Fig:ALT}   
	      	  \end{figure}

\begin{figure}
	        \includegraphics [height=10.5cm]{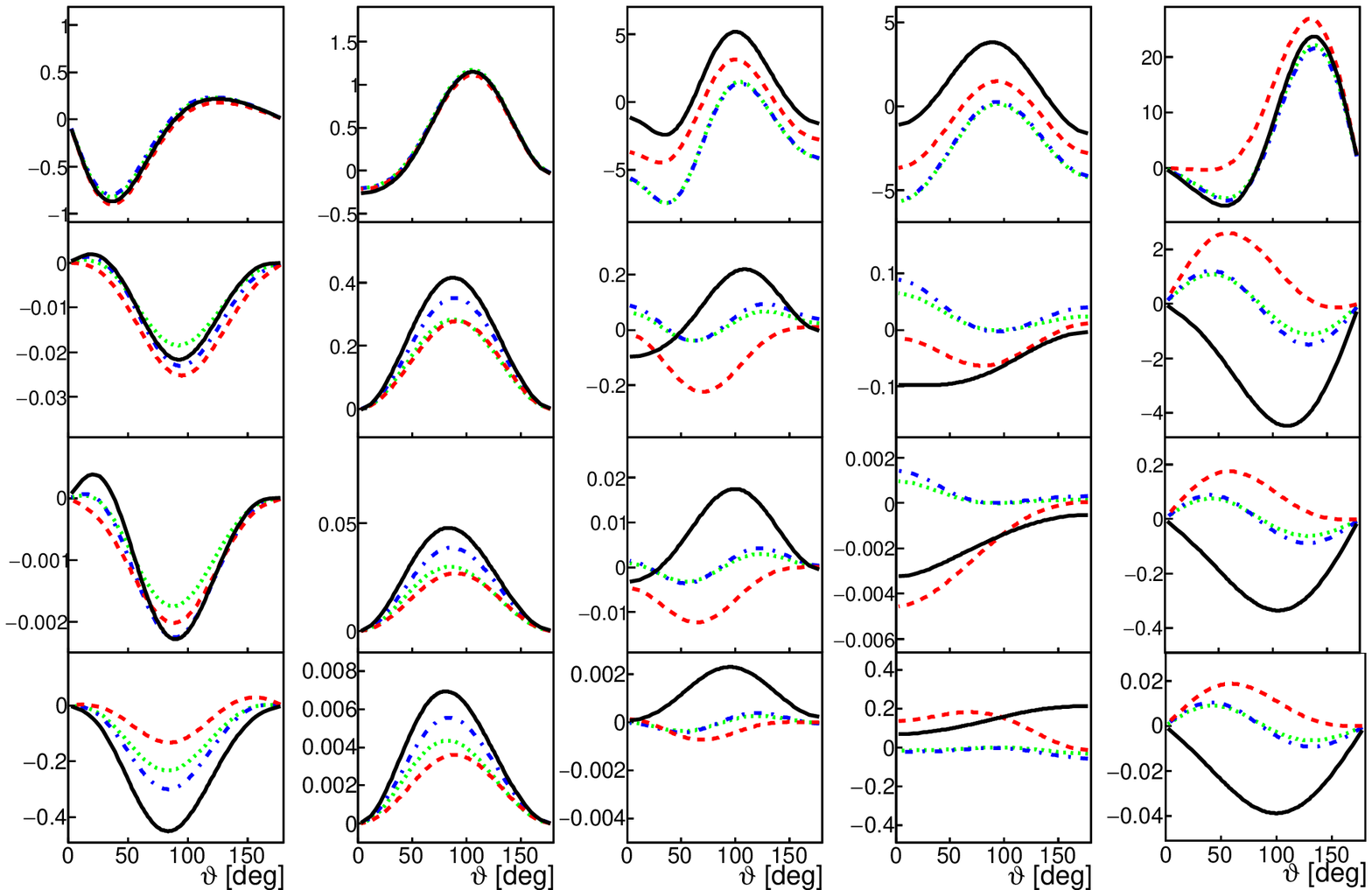}
	        \caption{$\vartheta$-dependence of the functions $B_{x}^{(TT)}$, $B_{z}^{(TT)}$, $B_{x}^{(LT)}$, $B_{y}^{(LT)}$, and $B_{z}^{(LT)}$ from left to right. Notations as in Fig. \protect\ref{Fig:HLT}. }
	         \label{Fig:BTT}   
	      	  \end{figure}

The asymmetries for a vector polarized target can be expressed as a function of longitudinal and transverse components, as in Eq. (\ref{eq:34a}). In Fig. \ref{Fig:ATT} the transverse components, i.e., the functions  $A_{x}^{(TT)}$, $A_{y}^{(TT)}$, $\bar{A}_{y}^{(TT)}$ and $A_{z}^{(TT)}$ (respectively from left to right) are illustrated in terms of the $\pi^0$ angle in the CMS of the $\gamma^*d$ system, and the longitudinal components, $A_{x}^{(LT)}$, $A_{y}^{(LT)}$, $A_{z}^{(LT)}$ and 
$A_{y}^{(LL)}$ are shown in Fig. \ref{Fig:ALT}. In Fig. \ref{Fig:BTT} the $\vartheta$-dependence of the functions $B_{x}^{(TT)}$, $B_{z}^{(TT)}$, $B_{x}^{(LT)}$ $B_{y}^{(LT)}$ and $B_{z}^{(LT)}$ is shown. 

		  
\begin{figure}
	        \includegraphics [height=10.5cm]{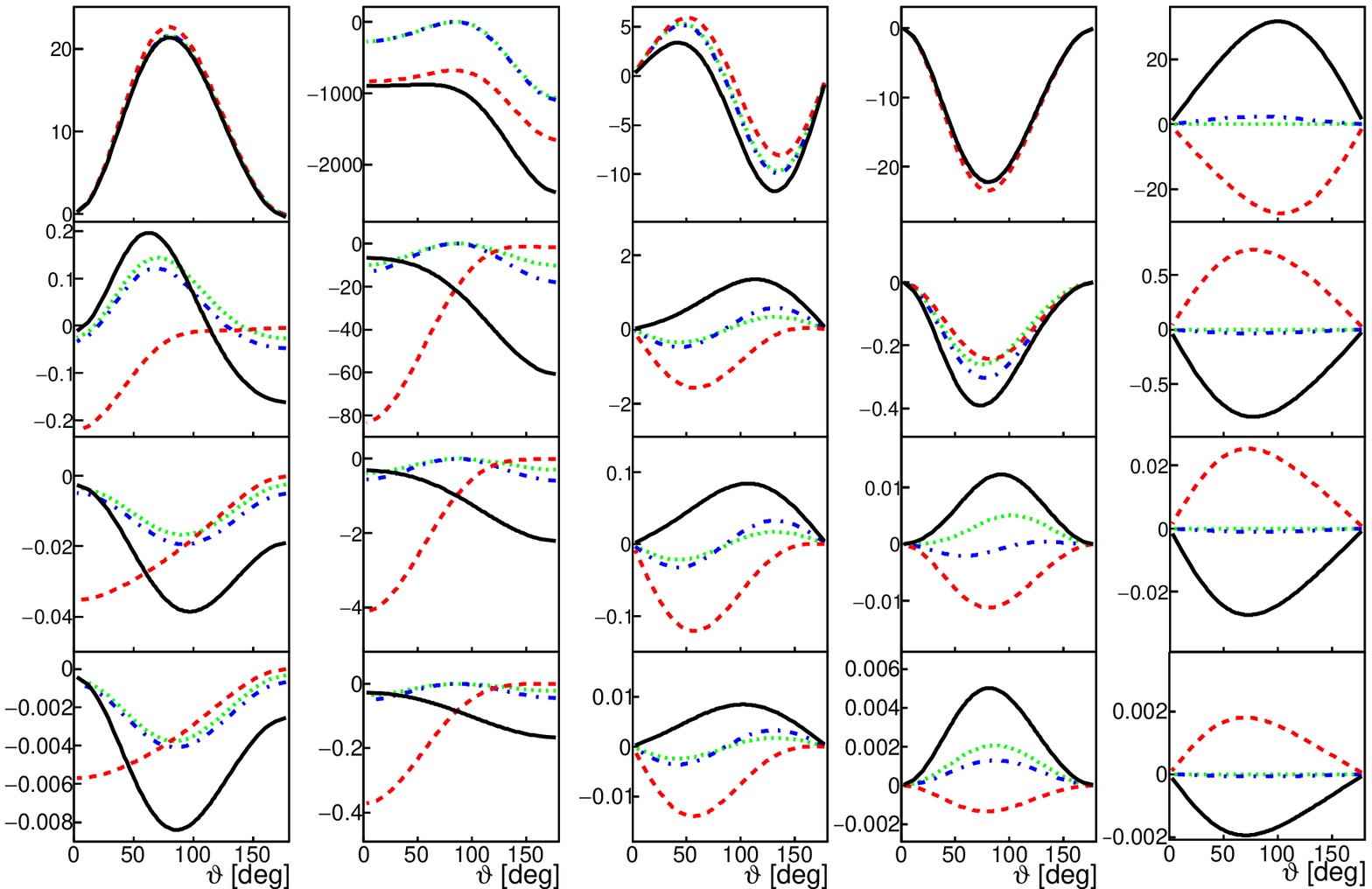}
	        \caption{$\vartheta$-dependence of the functions $A_{zz}^{T}$, $A_{zz}^{L}$, $A_{zz}^{I}$, $A_{zz}^{P}$ and $\bar A_{zz}^{I}$ from left to right. Notations as in Fig. \protect\ref{Fig:HLT}. }
	         \label{Fig:AzzIJ}   
	      	  \end{figure}

As an example of tensor observables, the $\vartheta$-dependence of the functions $A_{zz}^{T}$, $A_{zz}^{L}$, $A_{zz}^{I}$, $A_{zz}^{P}$ and $\bar A_{zz}^{I}$ (from left to right) is shown in  Fig. \protect\ref{Fig:AzzIJ}. 
 We can see that all observables have a strong angular dependence, they become smaller when $k^2$ increases, they may change shape and also the sign.
\section{Conclusions}
This work give general expressions for
various polarization observables in the coherent pseudoscalar meson
photo- and electroproduction on the deuteron target assuming one-photon-exchange approximation. 
It completes and generalizes a previous work on the unpolarized differential cross section.
The spin structure of the matrix element is explicitly derived in terms of structure functions. 
The correspondence with the helicity amplitudes is given.
The polarization
effects have been investigated for the case of a longitudinally
polarized electron beam and vector or tensor polarized deuteron target. In the case of the photoproduction
reaction, we consider linearly, circularly or elliptically polarized
photon beam. The asymmetries arising from the polarization of the particles in the initial state have been 
discussed as well as the measurable observables related to the scattered deuteron polarization.

Numerical
estimations  for the
unpolarized differential cross section and for some polarization
observables have been done in the frame of the simple model developed in Ref. \cite{Rekalo:2002km}.

The purpose of this paper is focussed on model independent expressions, that depend on the elementary amplitudes.
This formalism is applicable to all phenomenological models developed in frame of a definite picture of the deuteron,
and is useful to bridge the experimental information and the theoretical background.

\section{Acknowledgments}
This work was partially supported by the Ministry of Education and
Science of Ukraine (projects no. 0115U000474 and no. 0117U004866).
The research is carried on in the frame of the France-Ukraine IDEATE International Associated Laboratory (LIA).

\section{Appendix I: explicit expressions for the amplitudes $h$}
We present here the expressions for the structure functions $h_i$
and $\bar{h}_i$ (i=1-41) in terms of the scalar amplitudes  $g_i$
(i=1-13) describing the $\gamma ^*+d \to d+P$ reaction.

The structure functions $h_i$ describe the polarization observables
in the $\gamma ^*+d \to d+P$ reaction for the case of
different polarization states of the deuteron target.
\begin{itemize}
\item
\underline{The deuteron target is unpolarized.} The structure
functions $h_1-h_{5}$ corresponding to the interaction of the
virtual photon with an unpolarized deuteron target can be written as
\ba
h_1&=&\frac{1}{3}\biggl [|g_1|^2+\gamma_1^2|g_2|^2+a|g_3|^2+
b|g_4|^2+2cReg_3g_4^*\biggr ], \nn\\
h_2&=&\frac{1}{3}\biggl [|g_6|^2+a(|g_5|^2+\gamma_1^2|g_9|^2)+
b(|g_8|^2+\gamma_1^2|g_7|^2)+2cRe(g_5g_9^*+
\gamma_1^2g_7g_9^*)\biggr ], \nn\\
h_3&=&\frac{1}{3}\biggl [|g_{10}|^2+\gamma_1^2|g_{11}|^2+a|g_{12}|^2+
b|g_{13}|^2+2cReg_{12}g_{13}^*\biggr ], \nn\\
h_4&=&ReA_1, \ \ h_5=ImA_1, \nn\\
A_1&=&\frac{1}{3}\biggl [g_{1}g_{10}^*+\gamma_1^2g_{2}g_{11}^*+
ag_{3}g_{12}^*+bg_{4}g_{13}^*+c( g_{3}g_{13}^*+ g_{4}g_{12}^*)\biggr
], \nn\\
\gamma_1&=&\frac{E_1}{M}, \ \ a=1+\frac{{\vec q}^2}{M^2}\sin^2\vartheta , \ \
b=1+\frac{{\vec q}^2}{M^2} \cos^2\vartheta , \ \ c=\frac{{\vec
q}^2}{M^2} \cos\vartheta \sin\vartheta, 
\label{eq:A1}
\ea
where ${\vec q} (E_1)$ is
the $P-$meson momentum (energy) in the $\gamma ^*+d \to P+d$
reaction CMS and $\vartheta $ is the angle between the pseudoscalar
meson and virtual photon momenta in this system,
$E_1=(W^2-k^2+M^2)/2W.$
\item 

\underline{The deuteron target is a vector polarized.} The structure
functions $h_6-h_{18}$ which describe the effects of the vector
polarization of the deuteron target can be written as
\ba
h_6&=&-\frac{\gamma_1}{2}ImA_2, \ \ h_7=-\frac{\gamma_1}{2}ImA_3, \ \
h_8=\frac{\gamma_1}{2}ReA_2,  \ \ h_9=-\frac{\gamma_1}{2}ReA_3, \nn\\
A_2&=&g_{2}g_{6}^*-ag_{3}g_{9}^*-bg_{4}g_{7}^*-
c( g_{3}g_{7}^*+g_{4}g_{9}^*), 
\nn\\
A_3&=&-g_{6}g_{11}^*+ag_{9}g_{12}^*+bg_{7}g_{13}^*+
c( g_{9}g_{13}^*+g_{7}g_{12}^*), \nn\\
h_{10}&=&-\gamma_1Img_{1}g_{2}^*, ~
h_{11}=\gamma_1Im\biggl [-ag_{5}g_{9}^*+bg_{7}g_{8}^*-
c( g_{5}g_{7}^*+g_{8}g_{9}^*)\biggr ], \nn\\
h_{12}&=&-\gamma_1Img_{10}g_{11}^*,
h_{13}=-\frac{\gamma_1}{2}Im(g_{1}g_{11}^*-g_{2}g_{10}^*), ~
h_{14}=\frac{\gamma_1}{2}Re(g_{1}g_{11}^*-g_{2}g_{10}^*), \nn\\
h_{15}&=&-\frac{1}{2}ImA_4, \ \ h_{16}=-\frac{1}{2}ImA_5, 
h_{17}=\frac{1}{2}ReA_4,  \ \ h_{18}=-\frac{1}{2}ReA_5, \nn\\
A_4&=&-g_{1}g_{6}^*+ag_{3}g_{5}^*+bg_{4}g_{8}^*+
c( g_{3}g_{8}^*+g_{4}g_{5}^*), \nn\\
A_5&=&g_{6}g_{10}^*-ag_{5}g_{12}^*-bg_{8}g_{13}^*-
c( g_{5}g_{13}^*+g_{8}g_{12}^*). 
\label{eq:A2}
\ea

\item

\underline{The deuteron target is tensor polarized.} The structure
functions $h_{19}-h_{41}$ which describe the effects of the tensor
polarization of the deuteron target can be written as
\ba
h_{19}&=&|g_1|^2-\gamma_1^2|g_2|^2, ~
h_{20}=a|g_{5}|^2+b|g_{8}|^2+2cReg_{5}g_{8}^*-\nn\\
&&
\gamma_1^2\biggl [a|g_{9}|^2+b|g_{7}|^2+
2cReg_{7}g_{9}^*\biggr ], \nn\\
h_{21}&=&|g_{10}|^2-\gamma_1^2|g_{11}|^2, 
h_{22}=Re(g_{1}g_{10}^*-\gamma_1^2g_{2}g_{11}^*), \nn\\
h_{23}&=&Im(g_{1}g_{10}^*-\gamma_1^2g_{2}g_{11}^*), ~
h_{24}=a|g_{3}|^2+b|g_{4}|^2+2cReg_{3}g_{4}^*-
\gamma_1^2|g_{2}|^2, \nn\\
h_{25}&=&|g_{6}|^2-\gamma_1^2\biggl [a|g_{9}|^2+b|g_{7}|^2+
2cReg_{7}g_{9}^*\biggr ], \nn\\
h_{26}&=&a|g_{12}|^2+b|g_{13}|^2+2cReg_{12}g_{13}^*-
\gamma_1^2|g_{11}|^2, \nn\\
h_{27}&=&ReA_6, \ \ h_{28}=ImA_6, 
~A_6=ag_{3}g_{12}^*+bg_{4}g_{13}^*+c(g_{3}g_{13}^*+
g_{4}g_{12}^*)-\gamma_1^2g_{2}g_{11}^*, \nn\\
h_{29}&=&2Reg_{1}g_{2}^*, 
~h_{30}=2Re\biggl [ag_{5}g_{9}^*+bg_{7}g_{8}^*+
c(g_{5}g_{7}^*+g_{8}g_{9}^*)\biggr ], 
\nn\\
h_{31}&=&2Reg_{10}g_{11}^*, 
~h_{32}=Re(g_{2}g_{10}^*+g_{1}g_{11}^*), ~
h_{33}=Im(g_{2}g_{10}^*+g_{1}g_{11}^*), 
\nn\\
h_{34}&=&ReA_7, \ \ h_{35}=ReA_8, \ \
h_{36}=ImA_7,  \ \ h_{37}=-ImA_8, \nn\\
A_7&=&ag_{3}g_{5}^*+bg_{4}g_{8}^*+c(g_{3}g_{8}^*+
g_{4}g_{5}^*)+g_{1}g_{6}^*, \nn\\
A_8&=&g_{5}g_{12}^*+bg_{8}g_{13}^*+c(g_{5}g_{13}^*+
g_{8}g_{12}^*)+g_{6}g_{10}^*, \nn\\
h_{38}&=&ReA_9, \ \ h_{39}=ReA_{10}, \ \
h_{40}=ImA_9,  \ \ h_{41}=-ImA_{10}, \nn\\
A_9&=&ag_{3}g_{9}^*+bg_{4}g_{7}^*+c(g_{3}g_{7}^*+
g_{4}g_{9}^*)+g_{2}g_{6}^*, \nn\\
A_{10}&=&ag_{9}g_{12}^*+bg_{7}g_{13}^*+c(g_{9}g_{13}^*+
g_{7}g_{12}^*)+g_{6}g_{11}^*. 
\label{eq:A2a}
\ea

The structure functions $\bar{h}_i$ describe the polarization
observables in the $\gamma ^*+d \to d+P$ reaction for the
case of the polarized scattered deuteron.
\item

\underline{The scattered deuteron is vector polarized.} The
structure functions $\bar{h}_6-\bar{h}_{18}$ which describe the
effects of the vector polarization of the scattered deuteron can be
written as
\ba
\bar{h}_6&=&\frac{1}{6}[xIm(g_{4}g_{6}^*-g_{1}g_{8}^*-\gamma^2_1g_{2}g_{7}^*)-
zIm(g_{1}g_{5}^*-g_{3}g_{6}^*+\gamma^2_1g_{2}g_{9}^*), \nn\\
\bar{h}_7&=&-\frac{1}{6}[zIm(g_{6}g_{12}^*-g_{5}g_{10}^*-\gamma^2_1g_{9}g_{11}^*)-
xIm(g_{8}g_{10}^*-g_{6}g_{13}^*+\gamma^2_1g_{7}g_{11}^*), \nn\\
\bar{h}_8&=&\frac{1}{6}[-xRe(g_{4}g_{6}^*-g_{1}g_{8}^*-\gamma^2_1g_{2}g_{7}^*)+
zRe(g_{1}g_{5}^*-g_{3}g_{6}^*+\gamma^2_1g_{2}g_{9}^*), \nn\\
\bar{h}_9&=&-\frac{1}{6}[zRe(g_{6}g_{12}^*-g_{5}g_{10}^*-\gamma^2_1g_{9}g_{11}^*)-
xRe(g_{8}g_{10}^*-g_{6}g_{13}^*+\gamma^2_1g_{7}g_{11}^*), \nn\\
\bar{h}_{10}&=&\frac{1}{3}\gamma_2Img_{3}g_{4}^*, \ \
\bar{h}_{11}=\frac{1}{3}\gamma_2Im(g_{5}g_{8}^*-\gamma^2_1g_{7}g_{9}^*),
\ \ \bar{h}_{12}=\frac{1}{3}\gamma_2Img_{12}g_{13}^*,
\nn\\
\bar{h}_{13}&=&\frac{1}{6}\gamma_2Im(g_{3}g_{13}^*-g_{4}g_{12}^*), \ \
\bar{h}_{14}=-\frac{1}{6}\gamma_2Re(g_{3}g_{13}^*-g_{4}g_{12}^*),
\nn\\
\bar{h}_{15}&=&-\frac{1}{6}[zIm(g_{4}g_{6}^*-g_{1}g_{8}^*-\gamma^2_1g_{2}g_{7}^*)-
yIm(g_{1}g_{5}^*-g_{3}g_{6}^*+\gamma^2_1g_{2}g_{9}^*), \nn\\
\bar{h}_{16}&=&-\frac{1}{6}[zIm(g_{8}g_{0}^*-g_{6}g_{13}^*+\gamma^2_1g_{7}g_{11}^*)-
yIm(g_{6}g_{12}^*-g_{5}g_{10}^*-\gamma^2_1g_{9}g_{11}^*), \nn\\
\bar{h}_{17}&=&\frac{1}{6}[zRe(g_{4}g_{6}^*-g_{1}g_{8}^*-\gamma^2_1g_{2}g_{7}^*)-
yRe(g_{1}g_{5}^*-g_{3}g_{6}^*+\gamma^2_1g_{2}g_{9}^*), \nn\\
\bar{h}_{18}&=&-\frac{1}{6}[zRe(g_{8}g_{0}^*-g_{6}g_{13}^*+\gamma^2_1g_{7}g_{11}^*)-
yRe(g_{6}g_{12}^*-g_{5}g_{10}^*-\gamma^2_1g_{9}g_{11}^*), 
\label{eq:A3}
\ea
where
\be
x=\cos^2\vartheta\gamma_2+\sin^2\vartheta , \ \
y=\sin^2\vartheta\gamma_2+\cos^2\vartheta , \ \
z=(\gamma_2-1)\cos\vartheta \sin\vartheta , \ \
\gamma_2=\frac{E_2}{M}.
\nn
\ee
\item

\underline{The scattered deuteron is tensor polarized.} The
structure functions $\bar{h}_{19}-\bar{h}_{41}$ which describe the
effects of the tensor polarization of the scattered deuteron can be
written as
\ba
\bar{h}_{19}&=&\frac{1}{3d}(d|g_{3}|^2-u|g_{4}|^2), \nn\\
\bar{h}_{20}&=&\frac{1}{3d}[d(|g_{5}|^2+\gamma^2_1|g_{9}|^2)-
u(|g_{8}|^2+\gamma^2_1|g_{7}|^2)], \nn\\
\bar{h}_{21}&=&\frac{1}{3d}(d|g_{12}|^2-u|g_{13}|^2), \ \
\bar{h}_{22}=\frac{1}{3d}Re(dg_{3}g_{12}^*-ug_{4}g_{13}^*), \nn\\
\bar{h}_{23}&=&\frac{1}{3d}Im(dg_{3}g_{12}^*-ug_{4}g_{13}^*), \ \
\bar{h}_{24}=\frac{1}{3d}[d(|g_{1}|^2+\gamma^2_1|g_{2}|^2)-
\gamma^2_2|g_{4}|^2], \nn\\
\bar{h}_{25}&=&\frac{1}{3d}[d|g_{6}|^2-\gamma^2_2(|g_{8}|^2+\gamma^2_1|g_{7}|^2)], \nn\\
\bar{h}_{26}&=&\frac{1}{3d}[d(|g_{10}|^2+\gamma^2_1|g_{11}|^2)-
\gamma^2_2|g_{13}|^2], \nn\\
\bar{h}_{27}&=&\frac{1}{3d}Re[d(g_{1}g_{10}^*+\gamma^2_1g_{2}g_{11}^*)-
\gamma^2_2g_{4}g_{13}^*], \ \
\nn\\
\bar{h}_{28}&=&\frac{1}{3d}Im[d(g_{1}g_{10}^*+\gamma^2_1g_{2}g_{11}^*)-
\gamma^2_2g_{4}g_{13}^*], \nn\\
\bar{h}_{29}&=&\frac{2}{3d}[dReg_{3}g_{4}^*+(1+\gamma_2)z|g_{4}|^2],
\nn\\
\bar{h}_{30}&=&\frac{2}{3d}[d
Re(g_{5}g_{8}^*+\gamma^2_1g_{7}g_{9}^*)+(1+\gamma_2)z(|g_{8}|^2+\gamma^2_1|g_{7}|^2)],
\nn\\
\bar{h}_{31}&=&\frac{2}{3d}[dReg_{12}g_{13}^*+(1+\gamma_2)z|g_{13}|^2],
\nn\\
\bar{h}_{32}&=&\frac{1}{3d}Re[d(g_{4}g_{12}^*+g_{3}g_{13}^*)+2(1+\gamma_2)zg_{4}g_{13}^*],
\nn\\
\bar{h}_{33}&=&\frac{1}{3d}Im[d(g_{4}g_{12}^*+g_{3}g_{13}^*)+
2(1+\gamma_2)zg_{4}g_{13}^*], \nn\\
\bar{h}_{34}&=&\frac{1}{3}Re(g_{3}g_{6}^*+g_{1}g_{5}^*+\gamma^2_1g_{2}g_{9}^*),
\nn\\
\bar{h}_{35}&=&\frac{1}{3}Re(g_{6}g_{12}^*+g_{5}g_{10}^*+\gamma^2_1g_{9}g_{11}^*),
\ \
\bar{h}_{36}=\frac{1}{3}Im(g_{3}g_{6}^*+g_{1}g_{5}^*+\gamma^2_1g_{2}g_{9}^*),
\nn\\
\bar{h}_{37}&=&-\frac{1}{3}Im(g_{6}g_{12}^*+g_{5}g_{10}^*+\gamma^2_1g_{9}g_{11}^*),
\ \
\bar{h}_{38}=\frac{1}{3}Re(g_{4}g_{6}^*+g_{1}g_{8}^*+\gamma^2_1g_{2}g_{7}^*),
\nn\\
\bar{h}_{39}&=&\frac{1}{3}Re(g_{6}g_{13}^*+g_{8}g_{10}^*+\gamma^2_1g_{7}g_{11}^*),
\ \
\bar{h}_{40}=\frac{1}{3}Im(g_{4}g_{6}^*+g_{1}g_{8}^*+\gamma^2_1g_{2}g_{7}^*),
\nn\\
\bar{h}_{41}&=&-\frac{1}{3}Im(g_{6}g_{13}^*+g_{8}g_{10}^*+\gamma^2_1g_{7}g_{11}^*),
\label{eq:A4}
\ea
where $d=\cos^2\vartheta+\gamma^2_2\sin^2\vartheta  , \
u=\sin^2\vartheta+\gamma^2_2 \cos^2\vartheta . $

\end{itemize}


\end{document}